\documentclass[twoside,12pt]{article}

\usepackage{multicol}
\usepackage{amsmath}
\usepackage{amsfonts}
\usepackage{amssymb}
\usepackage{bm}
\usepackage{hyperref}
\usepackage{mathtools}

\renewcommand{\Re}{\mathrm{Re}}%

\numberwithin{equation}{section}

\begin{document}

\thispagestyle{empty}

\title{\bfseries Reply to \\
  ``Extended Rejoinder to \\ 
   ``Extended Comment on \\
  ``One-Range Addition Theorems for \\ 
  Coulomb Interaction Potential \\ 
  and Its Derivatives'' \\
  by I.\ I.\ Guseinov \\
  (Chem.\ Phys.\ Vol.\ 309 (2005), \\ 
  pp.\ 209 - 211)'', arXiv:0706.0975v2''}

\author{Ernst Joachim Weniger \\
Institut f\"ur Physikalische und Theoretische Chemie \\
Universit\"at Regensburg, D-93040 Regensburg, Germany}

\date{Submitted to the Los Alamos Preprint Server: 23 July 2007}

\maketitle

\begin{abstract}
  \noindent
  In the years from 2001 to 2006, Guseinov and his coworkers published 40
  articles on the derivation and application of one-range addition
  theorems. In E.~J.\ Weniger, Extended Comment on ``One-Range Addition
  Theorems for Coulomb Interaction Potential and Its Derivatives'' by I.\
  I.\ Guseinov (Chem. Phys. Vol. 309 (2005), pp. 209 - 213),
  arXiv:0704.1088v2 [math-ph], it was argued that Guseinov's treatment of
  one-range addition theorems is at best questionable and in some cases
  fundamentally flawed. In I.\ I.\ Guseinov, Extended Rejoinder to
  ``Extended Comment on "One-Range Addition Theorems for Coulomb
  Interaction Potential and Its Derivatives'' by I. I. Guseinov (Chem.
  Phys. and Vol. 309 (2005)'', pp. 209-213), arXiv:0706.0975v2
  [physics.chem-ph], these claims were disputed. To clarify the
  situation, the most serious mathematical flaws in Guseinov's treatment
  of one-range addition theorems are discussed in more depth.
\end{abstract}

\newpage

\tableofcontents

\newpage

\typeout{==> Section: Introduction}
\section{Introduction}
\label{Sec:Intro}

The efficient and reliable evaluation of multicenter integrals is among
the oldest mathematical and computational problems of molecular
electronic structure theory. In spite of heroic efforts, the mathematical
and computational problems, that occur in this context, are not yet
solved in a completely satisfactory way, and there is still a
considerable amount of research going on. Particularly difficult are
multicenter integrals of the physically better motivated exponentially
decaying functions, whose efficient and reliable evaluation is -- in
spite of all the mathematical and computational advances of recent years
-- still very difficult.

From a methodological point of view, research on multicenter integrals is
essentially mathematical in nature, although relatively few
mathematicians have been involved in this research. In my opinion, this
is quite deplorable. But within mathematics, research on multicenter
integrals is highly interdisciplinary. Absolutely essential is a good
knowledge of special function theory and of classical analysis: The
derivation of explicit expressions for multicenter integrals is to a
large extent some kind of 19th century mathematics. However, in order to
succeed we also need modern mathematical concepts as for example Hilbert
spaces, approximation theory, generalized functions, and angular momentum
theory.

It is the ultimate goal of research on multicenter integrals to produce
computer code that permits an efficient and reliable evaluation of these
integrals. Accordingly, a good knowledge of sophisticated numerical
techniques is absolutely indispensable.

Research on multicenter integrals is difficult, and there are many
chances of making errors. Firstly, there are errors that violate basic
mathematical principles, which could be called \emph{first-order errors}.
Secondly, there are mathematically correct manipulations and/or
deductions, which lead to computer code that is either hopelessly
inefficient or unreliable and which could be called \emph{second-order
  errors}.

Multicenter integrals are difficult to evaluate because the integration
variables occur in unseparated form. Principal mathematical tools, that
can accomplish such a separation of variables, are so-called addition
theorems. These are expansions of a given function $f (\bm{r} \pm
\bm{r}')$ with $\bm{r}, \bm{r}' \in \mathbb{R}^3$ in products of other
functions that only depend on either $\bm{r}$ or $\bm{r}'$.

In the articles
\cite{Guseinov/2001,Guseinov/2002b,Guseinov/2002c,Guseinov/2002d,%
  Guseinov/2003b,Guseinov/2003c,Guseinov/2003d,Guseinov/2003e,%
  Guseinov/2004a,Guseinov/2004b,Guseinov/2004c,Guseinov/2004d,%
  Guseinov/2004e,Guseinov/2004f,Guseinov/2004i,Guseinov/2004k,%
  Guseinov/2005a,Guseinov/2005b,Guseinov/2005c,Guseinov/2005d,%
  Guseinov/2005e,Guseinov/2005f,Guseinov/2005g,Guseinov/2006a,%
  Guseinov/2006b,Guseinov/Aydin/Mamedov/2003,Guseinov/Mamedov/2001c,%
  Guseinov/Mamedov/2002d,Guseinov/Mamedov/2003,Guseinov/Mamedov/2004b,%
  Guseinov/Mamedov/2004d,Guseinov/Mamedov/2004e,Guseinov/Mamedov/2004g,%
  Guseinov/Mamedov/2004h,Guseinov/Mamedov/2005c,Guseinov/Mamedov/2005d,%
  Guseinov/Mamedov/2005g,Guseinov/Mamedov/Rzaeva/2001,%
  Guseinov/Mamedov/Rzaeva/2002,Guseinov/Mamedov/Suenel/2002}, Guseinov
and coworkers derived so-called one-range addition theorems for a
variety of different functions and applied them for the evaluation of
multicenter integrals.

In my Comment \cite{Weniger/2007b}, I presented a very detailed criticism
of Guseinov's work on one-range addition theorems and showed that both
first- and second-order errors occur quite abundantly. I also criticized
in \cite{Weniger/2007b} that Guseinov's articles are often remarkably
similar and that they do not always give due credit to the previous work
of others. Moreover, I made several suggestions how Guseinov's flawed
approach could be improved: In \cite[Section 7]{Weniger/2007b} I
suggested to employ advanced mathematical concepts from the theory of
generalized functions in order to give a meaning to divergent one-range
addition theorems, or to use nonlinear sequence transformations (see for
example \cite{Weniger/1989,Weniger/2004} and references therein), either
to sum divergent series, whose occurrence was apparently either
overlooked or ignored, or to accelerate the convergence of series
expansions for multicenter integrals, whose convergence need not be
rapid.

Apparently, my Comment \cite{Weniger/2007b} did not impress Guseinov too
much, who recently wrote a Rejoinder \cite{Guseinov/2007a} to my Comment
\cite{Weniger/2007b} In his Abstract, Guseinov stated
\begin{quote}
  \sl The concrete criticism raised in Weniger's comment against our
  papers actually touches a very minor aspect of the works that are not
  relevant at all for the conclusions, which are made.
\end{quote}
and
\begin{quote}
  \sl All claims of inconsistencies and flaws in the theoretical
  framework are rejected as unfounded. This rejoinder paper contains all
  of the answers to Weniger's comments.
\end{quote}
This is wrong. In \cite{Guseinov/2007a}, Guseinov did not address at all
the most serious and consequential first-order errors, which I had
criticized in my Comment \cite{Weniger/2007b} and which raise serious
doubts on the mathematical soundness of Guseinov's treatment of one-range
addition theorems. Guseinov's questionable attitude towards mathematical
rigor becomes also evident in his even more recent preprints
\cite{Guseinov/2007b,Guseinov/2007c}, in which he proceeds in the same
style and spirit as in his previous articles on one-range addition
theorems mentioned above, completely ignoring my criticism.

In this Replay to Guseinov's Rejoinder \cite{Guseinov/2007a}, I discuss
once more and in more depth the most important mathematical flaws of
Guseinov's work. In contrast to my earlier and longer Comment
\cite{Weniger/2007b}, I concentrate entirely on first-order errors.

In \cite{Guseinov/2001,Guseinov/2002c}, Guseinov had derived one-range
addition theorems for Slater-type functions
\begin{equation}
  \label{Def_STF}
  \chi_{N, L}^{M} (\beta, \mathbf{r}) \; = \;
  (\beta r)^{N-L-1} \, \mathrm{e}^{- \beta r} \,
  \mathcal{Y}_{L}^{M} (\beta \bm{r}) \, , \qquad \beta > 0 \, ,
\end{equation}
with in general nonintegral principal quantum numbers $N \in \mathbb{R}
\setminus \mathbb{N}$ by expanding them in terms his functions
$\prescript{}{k}{\Psi}_{n, \ell}^{m} (\beta, \bm{r})$ defined by
(\ref{Def_Psi_Guseinov}), which are complete and orthonormal in the
weighted Hilbert spaces $L_{r^k}^{2} (\mathbb{R}^3)$ defined by
(\ref{HilbertL_r^k^2}) (for details, see \cite[Sections 4 -
6]{Weniger/2007b}). Here, $\mathcal{Y}_{L}^{M} (\beta \bm{r})$ is a
regular solid harmonic (compare \cite[Eq.\ (A.2)]{Weniger/2007b})).

As long as the principal quantum numbers $N$ are not too negative,
Slater-type functions $\chi_{N, L}^{M}$ belong for $k=-1, 0, 1, 2, \dots$
to the weighted Hilbert spaces $L_{r^k}^{2} (\mathbb{R}^3)$ defined by
(\ref{HilbertL_r^k^2}), which was implicitly used by Guseinov.  In this
case, the one-range addition theorems derived in
\cite{Guseinov/2001,Guseinov/2002c} exist and converge in the mean with
respect to the norms (\ref{Norm_r^k_2}) of these Hilbert spaces.

Guseinov's addition theorems for Slater-type functions $\chi_{N, L}^{M}$
yield for $N=L=M=0$ the corresponding addition theorems for the Yukawa
potential $\exp (- \beta r)/r$. Guseinov \cite{Guseinov/2005a} derived
his one-range addition theorems for the Coulomb potential by exploiting
the obvious relationship $1/r = \lim_{\beta \to 0} \exp (- \beta r)/r$ in
his one-range addition theorems for the Yukawa potential.

At first sight, it may look like a good idea to derive one-range addition
theorems for the Coulomb potential by performing the comparatively simple
limit $\beta \to 0$ in the one-range addition theorems for the Yukawa
potential. This pragmatic approach seems to permit an efficient
utilization of the available information and of previous work.
Unfortunately, the situation is much more complicated.

As shown in Section \ref{Sec:ContLim}, the limit $\beta \to 0$ in
mathematically well defined integrals containing the Yukawa potential is
not necessarily continuous and does not always lead to a finite result.
So, whenever we try to find an expression for an integral containing the
Coulomb potential as the limiting case of the expression for an analogous
integral containing the Yukawa potential, we have to be cautious and be
prepared for complications. As discussed in Sections \ref{Sec:DivOrthExp}
and \ref{Sec:OneRangAddThmCP}, similar complications can occur also in
orthogonal expansions of the Yukawa potential.

A very serious weakness of Guseinov's earlier work on one-range addition
theorems is that he completely ignored the obvious fact that orthogonal
expansions can diverge, and he still does this in his recent Rejoinder
\cite{Guseinov/2007a} and in his even more recent preprints
\cite{Guseinov/2007b,Guseinov/2007c}. On p.\ 3 of his Rejoinder
\cite{Guseinov/2007a}, Guseinov states:
\begin{quote}
  \sl The essential facts of Hilbert space and approximation theory as
  well as all questions of convergence and existence have been taken into
  account by Guseinov and his coworkers in the context of one-range
  addition theorems and multicenter integrals. Thus, the Guseinov's
  treatment of one-range addition theorems is not questionable, and is
  fundamentally flawless from a mathematical point of view.
\end{quote}
Again, this is wrong. Let us assume that $\mathcal{H}$ is a Hilbert
space, and that $\{ \varphi_n \}_{n=0}^{\infty}$ is a function set that
is complete and orthonormal in $\mathcal{H}$. If $f \in \mathcal{H}$,
then standard Hilbert space theory tells us that the expansion
$\sum_{n=0}^{\infty} (\varphi_n \vert f) \varphi_n$ converges to $f$ in
the mean with respect to the norm of $\mathcal{H}$.  If, however, $f
\notin \mathcal{H}$, then it follows -- as shown in Section
\ref{Sec:DivOrthExp} -- from the Riesz-Fischer Theorem (see for example
\cite[Theorem 7.43 on p.\ 191]{Griffel/2002}) that the formal orthogonal
expansion $f = \sum_{n=0}^{\infty} (\varphi_n \vert f) \varphi_n$
diverges in the mean with respect to the norm of $\mathcal{H}$.

Since Guseinov ignores the fact that orthogonal expansions can diverge,
he fails to take into account that divergent orthogonal expansions cannot
be treated like convergent orthogonal expansions (obviously, the
divergence of an orthogonal series can do a lot of harm in integrals).
So, Guseinov never wonders whether and under which conditions divergent
expansions can safely be used in multicenter integrals, although this is
by no means obvious.

As shown in Section \ref{Sec:DivOrthExp}, it is nevertheless possible to
use divergent orthogonal expansions in inner products or other
functionals -- in our case usually multicenter integrals -- in a
mathematically meaningful way. The key is that the functionals, in which
these expansions are to be used, have to satisfy additional and possibly
very restrictive regularity conditions. This follows from the Riesz
Representation Theorem (see for example \cite[Theorem 7.60 on p.\
199]{Griffel/2002}). Accordingly, divergent orthogonal expansions are
essentially generalized functions in the sense of Schwartz
\cite{Schwartz/1966a} that can -- in spite of their divergence --
converge weakly when used in suitably restricted functionals.

When Guseinov derived in \cite{Guseinov/2005a} one-range addition
theorems for the Coulomb potential by considering the limit $\beta \to 0$
in one-range addition theorems for the Yukawa potential $\exp (-\beta
r)/r$, he overlooked some very consequential facts.

The Yukawa potential belongs for $k=0, 1, 2, \dots$, to the weighted
Hilbert space $L_{r^k}^{2} (\mathbb{R}^3)$1 defined by
(\ref{HilbertL_r^k^2}), but not for $k=-1$ (this obvious fact was
apparently overlooked by Guseinov). Accordingly, Guseinov's one-range
addition theorems for the Yukawa potential converge for $k = 0, 1, 2,
\dots$ in the mean with respect to the norm (\ref{Norm_r^k_2}) of
$L_{r^k}^{2} (\mathbb{R}^3)$, but not for $k=-1$ (see also Appendix
\ref{App:ExpYukawa2GusFun} where the convergence of the one-center limit
of these addition theorems is analyzed).

As discussed in Section \ref{Sec:OneRangAddThmCP}, the Coulomb potential
does not belong to any of the Hilbert spaces $L_{r^k}^{2} (\mathbb{R}^3)$
which Guseinov had implicitly used. This is quite consequential:
Guseinov's one-range addition theorems for the Coulomb potential diverge
for all $k=-1, 0, 1, 2, \dots$ in the mean with respect to the norm
(\ref{Norm_r^k_2}) of $L_{r^k}^{2} (\mathbb{R}^3)$, although they were
derived by a limiting procedure from the corresponding addition theorems
of the Yukawa potential, which converge at least for $k=0, 1, 2, \dots$
in the mean. This is another example that the limit $\beta \to 0$ in
expressions involving the Yukawa potential $\exp (-\beta r)/r$ need not
be continuous and does not necessarily produce something finite.

Section \ref{Sec:OneRangAddThmCP} shows conclusively that it is
impossible to construct one-range addition theorems for the Coulomb
potential that converge in the mean with respect to the norms of the
weighted Hilbert spaces $L_{r^k}^{2} (\mathbb{R}^3)$. More advanced
mathematical concepts such as the theory of generalized functions or
possibly also powerful numerical techniques for the summation of
divergent series (see for example \cite{Weniger/1989,Weniger/2004} and
references therein) are needed to give divergent orthogonal expansions of
that kind any meaning beyond purely formal expansions. None of these
things are discussed in Guseinov's articles on one-range addition
theorems
\cite{Guseinov/2001,Guseinov/2002b,Guseinov/2002c,Guseinov/2002d,%
  Guseinov/2003b,Guseinov/2003c,Guseinov/2003d,Guseinov/2003e,%
  Guseinov/2004a,Guseinov/2004b,Guseinov/2004c,Guseinov/2004d,%
  Guseinov/2004e,Guseinov/2004f,Guseinov/2004i,Guseinov/2004k,%
  Guseinov/2005a,Guseinov/2005b,Guseinov/2005c,Guseinov/2005d,%
  Guseinov/2005e,Guseinov/2005f,Guseinov/2005g,Guseinov/2006a,%
  Guseinov/2006b,Guseinov/Aydin/Mamedov/2003,Guseinov/Mamedov/2001c,%
  Guseinov/Mamedov/2002d,Guseinov/Mamedov/2003,Guseinov/Mamedov/2004b,%
  Guseinov/Mamedov/2004d,Guseinov/Mamedov/2004e,Guseinov/Mamedov/2004g,%
  Guseinov/Mamedov/2004h,Guseinov/Mamedov/2005c,Guseinov/Mamedov/2005d,%
  Guseinov/Mamedov/2005g,Guseinov/Mamedov/Rzaeva/2001,%
  Guseinov/Mamedov/Rzaeva/2002,Guseinov/Mamedov/Suenel/2002} or in his
recent Rejoinder \cite{Guseinov/2007a}.

Unfortunately, this is not yet the end of Guseinov's grave first-order
errors in his treatment of one-range addition theorems. As discussed in
Section \ref{Sec:RearrAddThm}, Guseinov preferred to replace the complete
and orthonormal expansion functions $\prescript{}{k}{\Psi}_{n, \ell}^{m}
(\beta, \bm{r})$ of his one-range addition theorems for Slater-type
functions by nonorthogonal Slater-type functions with integral principal
quantum numbers and to rearrange the order of summations of the resulting
expansions. In this way, Guseinov obtained expansions of Slater-type
functions $\chi_{N, L}^{M} (\beta, \mathbf{r} \pm \mathbf{r}')$ with in
general nonintegral principal quantum numbers $N \in \mathbb{R} \setminus
\mathbb{N}$ in terms of Slater-type functions $\chi_{n, \ell}^{m} (\beta,
\mathbf{r})$ with integral principal quantum numbers $n \in \mathbb{N}$
located at a different center.

As is well known, Slater-type functions are complete in the Hilbert
spaces implicitly used by Guseinov, but not orthogonal. This is very
consequential. It is extensively documented both in the mathematical
literature (see for example \cite[Theorem 10 on p.\ 54]{Davis/1989} or
\cite[Section 1.4]{Higgins/1977}) as well as in the literature on
electronic structure calculations
\cite{Klahn/1975,Klahn/Bingel/1977a,Klahn/Bingel/1977b,%
  Klahn/Bingel/1977c,Klahn/1981,Klahn/Morgan/1984}) that the existence of
expansions in terms of nonorthogonal function sets is not guaranteed in
the case of essentially arbitrary functions. Such an expansions may or
may not exist. Thus, as already emphasized in \cite[Section
6]{Weniger/2007b}), Guseinov's approach is dangerous and potentially
disastrous and the validity of his rearranged addition theorems has to be
checked explicitly.

As discussed in Section \ref{Sec:RearrAddThm}, Guseinov disagreed in his
Rejoinder \cite[p.\ 7]{Guseinov/2007a} with my conclusions, and claimed
instead that the validity of his approach follows from Eq.\ (3.11) of his
Rejoinder. It is easy to show that Guseinov's reasoning is superficial
and that it is indeed necessary to analyze whether Guseinov's rearranged
addition theorems exist or not.

There is the practical problem that one-range additions theorems for
exponentially decaying functions are fairly complicated mathematical
objects. Accordingly, explicit proofs of their convergence or divergence
are very difficult and would most likely require a considerable amount of
time and effort. Fortunately, at least some insight can be gained by
analyzing instead the much simpler one-center limits of Guseinov's
one-range addition theorems, although this approach does not answer all
questions of interest.

In Section \ref{Sec:RearrAddThm} and also in \cite[Section
6]{Weniger/2007b}, it is shown that it is impossible to rearrange the
one-center limits of Guseinov's one-range addition theorems for
Slater-type functions if the principal quantum number $N$ is nonintegral,
$N \in \mathbb{R} \setminus \mathbb{N}$. Guseinov's rearrangements do not
lead to divergent series in the usual sense, but to power series with
series coefficients that are for all but a finite number of indices
\emph{infinite}. While I can probably claim with some confidence that I
have a lot of experience with the summation of divergent series (see for
example \cite{Weniger/1989,Weniger/2004,Weniger/2007a} and references
therein), I nevertheless must admit that I have not the slightest idea
what to do with power series with an infinite number of infinite terms.

This article is concluded by a Summary in Section (\ref{Sec:SumConclu}).
For the convenience of the readers, the most important conventions and
definitions of this Reply are listed in Appendices (\ref{App:Terminolgy})
- (\ref{App:GusFun}). Finally, there is Appendix
(\ref{App:ExpYukawa2GusFun}) analyzing the convergence of the one-center
limit of Guseinov's one-range addition theorems for the Yukawa potential
in the weighted Hilbert space $L_{r^k}^{2} (\mathbb{R}^3)$.

\typeout{==> Section: On the Continuity of Limits in Integrals}
\section{On the Continuity of Limits in Integrals}
\label{Sec:ContLim}

Because of its exponential decay, the Yukawa potential $\exp (- \beta
r)/r$ is in many respects a much more convenient mathematical object than
the closely related Coulomb potential $1/r$. This is particularly true
for integrals over the whole three-dimensional space $\mathbb{R}^{3}$ as
they occur in atomic or molecular electronic structure calculations.

Consequently, it is an obvious idea to derive explicit expressions for
integrals involving the Coulomb potential by performing the limit $\beta
\to 0$ in explicit expressions for analogous integrals involving the more
convenient Yukawa potential.

Often, this indirect approach is very effective. However, it is no
panacea. Moreover, it can easily lead to problems: The limiting process
$1/r = \lim_{\beta \to 0} \exp (- \beta r)/r$ is not necessarily
continuous in integrals, and it is not guaranteed that it produces a
finite result.

These possible problems can be illuminated easily by considering
integrals of the following kind:
\begin{equation}
  \label{Def_J_Int}
  \mathcal{J} (f; \beta) \; = \;
  \int \, \bigl[ f (\bm{r}) \bigr]^{*} \, \frac{\exp (-\beta r)}{r} \,
  \mathrm{d}^3 \bm{r}
\end{equation}
As usual, integration extends over the whole $\mathbb{R}^{3}$.

For $\beta > 0$, the Yukawa potential belongs to the Hilbert space $L^{2}
(\mathbb{R}^3)$ of square integrable functions defined by
(\ref{HilbertL^2}). If we also have $f \in L^{2} (\mathbb{R}^3)$, the
integral (\ref{Def_J_Int}) is a special case of the inner products
(\ref{InnerProd}), and it is finite. This follows at once form the
\emph{Cauchy-Schwarz Inequality} (see for example \cite[Theorem 7.7 on
p.\ 177]{Griffel/2002}) which can be expressed as follows:
\begin{equation}
  \label{CauchyScwarz}
  \vert (f \vert g) \vert^{2} \; \leq \; (f \vert f) \, (g \vert g) 
  \; = \; \Vert f \Vert^{2} \, \Vert g \Vert^{2} \, .
\end{equation}

If we perform the limit $\beta \to 0$ in (\ref{Def_J_Int}), we formally
obtain integrals of the following kind:
\begin{equation}
  \label{Def_K_Int}
  \mathcal{K} (f) \; = \;
  \int \, \bigl[ f (\bm{r}) \bigr]^{*} \, \frac{1}{r} \,
  \mathrm{d}^3 \bm{r}
\end{equation}
Since the Coulomb potential does not belong to $L^{2} (\mathbb{R}^3)$ or
to any of those Hilbert spaces, which are considered in this Reply and
which all involve an integration over the whole $\mathbb{R}^{3}$, the
Cauchy-Schwarz Inequality cannot be used to guarantee the existence of
$\mathcal{K} (f)$ for arbitrary $f \in L^{2} (\mathbb{R}^3)$. Instead,
the integral $\mathcal{K} (f)$ makes sense only if $f$ belongs to a
suitably restricted (proper) subset of $L^{2} (\mathbb{R}^3)$ pr to other
function spaces.

The possible discontinuity of the limit $\beta \to 0$ also becomes
evident in the six-dimensional integrals
\begin{equation}
  \label{CouInt_f_g}
  \mathcal{C} (f, g) \; = \; \int \! \int \, \bigl[ f (\bm{r}) \bigr]^{*} \,
  \frac{1}{\vert \bm{r} - \bm{r}' \vert} \, g (\bm{r}') \,
  \mathrm{d} \bm{r} \, \mathrm{d} \bm{r}'
\end{equation}
and
\begin{equation}
  \label{YukawaInt_f_g}
  \mathcal{Y} (f, g; \beta) \; = \; 
  \int \! \int \, \bigl[ f (\bm{r}) \bigr]^{*} \, 
  \frac {\exp (- \beta \vert \bm{r} - \bm{r}' \vert)}
  {\vert \bm{r} - \bm{r}' \vert} \, g (\bm{r}') \,
  \mathrm{d} \bm{r} \, \mathrm{d} \bm{r}' \, ,
\end{equation}
which describe the interaction of two charge densities $f, g\colon
\mathbb{R}^{3} \to \mathbb{C}$ via the Coulomb and the Yukawa potential,
respectively, and which involve an integration over the whole
six-dimensional space $\mathbb{R}^{3} \times \mathbb{R}^{3}$.

Obviously, we have $\mathcal{C} (f, g) = \lim_{\beta \to 0} \mathcal{Y}
(f, g; \beta)$, but it would be grossly negligent to perform this limit
without explicitly knowing criteria, which the charge densities $f$ and
$g$ have to satisfy in order to guarantee that the limit $\beta \to 0$
is continuous and produces a finite result.

This question can be analyzed with the help of Fourier transformation. If
we use the symmetrical version of Fourier transformation according to
(\ref{Def_FT}) and (\ref{Def_InvFT}), the six-dimensional integrals
(\ref{CouInt_f_g}) and (\ref{YukawaInt_f_g}) can be expressed as
three-dimensional Fourier integrals via the following general expression
introduced into electronic structure calculations by Geller \cite[Eqs.\
(1) and (2)]{Geller/1963b}:
\begin{equation}
  \label{6dimIntCR->3dimIntMR}
  \int \! \int \, f^{*} (\bm{r}) \, g (\bm{r}') \, h (\bm{r} - \bm{r}') \,
  \mathrm{d} \bm{r} \, \mathrm{d} \bm{r}' 
  \; = \; (2\pi)^{3/2} \, 
  \int \, \bigl[ \bar{f} (\bm{p}) \bigr]^{*} \, \bar{g} (\bm{p}) 
  \, \bar{h} (\bm{p}) \, \mathrm{d}^3 \bm{p} \, .
\end{equation}
If we choose $h (\bm{r} - \bm{r}') = \exp (-\beta [\bm{r} -
\bm{r}'])/\vert \bm{r} - \bm{r}' \vert$, we only need the Fourier
transform of the Yukawa potential (see for example \cite[Eqs.\ (6.8) and
(6.9)]{Weniger/2005}),
\begin{equation}
  \label{FT_YukPot}
  (2\pi)^{-3/2} \int \, \frac{\exp (-\mathrm{i} \bm{p} \cdot
    \bm{r}) \, \exp (-\beta r)}{r} \, \mathrm{d}^3 \bm{r} 
  \; = \; \frac{(2/\pi)^{1/2}}{\beta^{2} +  p^{2}} \, ,
  \qquad \beta > 0 \, ,
\end{equation}
to obtain:
\begin{equation}
  \label{YukawaInt_f_g_FT}
  \mathcal{Y} (f, g; \beta) \; = \; 4\pi \, 
  \int \, \bigl[ \bar{f} (\bm{p}) \bigr]^{*} \, \frac{1}{\beta^{2}+p^{2}}
  \, \bar{g} (\bm{p}) \, \mathrm{d}^3 \bm{p} \, .
\end{equation}
Fourier transformation maps $L^{2} (\mathbb{R}^3)$ onto $L^{2}
(\mathbb{R}^3)$ in a one-to-one manner such that inner products are
conserved \cite[Theorem IX.6 on p.\ 10]{Reed/Simon/1975}. Thus, $u, v \in
L^{2} (\mathbb{R}^3)$ implies $\bar{u}, \bar{v} \in L^{2}
(\mathbb{R}^3)$, and the coordinate and momentum space inner products
(\ref{InnerProd}) and (\ref{InnerProd_FT}), respectively, are identical:
\begin{equation}
  \label{InProdCons}
  \int \, \bigl[ u (\bm{r}) \bigr]^{*} \, v (\bm{r}) \,
  \mathrm{d}^3 \bm{r} \; = \; \int \, \bigl[ \bar{u} (\bm{p}) \bigr]^{*} \,
  \bar{v} (\bm{p})  \, \mathrm{d}^3 \bm{p} 
\end{equation}

It follows at once from the \emph{Cauchy-Schwarz Inequality}
(\ref{CauchyScwarz}) that $\mathcal{Y} (f, g; \beta)$ is finite if
$\bar{f} (\bm{p})/[\beta^{2}+p^{2}]^{1/2}, \bar{g}
(\bm{p})/[\beta^{2}+p^{2}]^{1/2} \in L^{2} (\mathbb{R}^3)$.
Unfortunately, this condition is not particularly helpful in practice
since it is not so easy to find convenient coordinate representations for
the functions $\bar{f} (\bm{p})/[\beta^{2}+p^{2}]^{1/2}$ and $\bar{g}
(\bm{p})/[\beta^{2}+p^{2}]^{1/2}$ (they have to be expressed as
convolution integrals containing $f (\bm{r})$ and $g (\bm{r})$,
respectively, multiplied by a modified Bessel function).

Fortunately, a slightly more restrictive, but much more convenient
condition on the charge densities $f$ and $g$ can be constructed easily.
For all $\bm{p} \in \mathbb{R}^{3}$ and for all $\beta > 0$, we have
$1/(\beta^{2}+p^{2}) \leq 1/\beta^{2}$. Thus,
\begin{equation}
  \left\vert \int \, \bigl[ \bar{f} (\bm{p}) \bigr]^{*}
    \, \frac{1}{\beta^{2}+p^{2}} \, \bar{g} (\bm{p}) \, \mathrm{d}^3 \bm{p}
  \right\vert  \; \leq \; \frac{1}{\beta^{2}} \, 
  \left\vert \int \, \bigl[ \bar{f} (\bm{p}) \bigr]^{*}
    \, \bar{g} (\bm{p}) \, \mathrm{d}^3 \bm{p}
  \right\vert \, , \qquad \beta > 0 \, .
\end{equation}
By applying the Cauchy-Schwarz Inequality (\ref{CauchyScwarz}) to the
integral on the right-hand side we find that $\mathcal{Y} (f, g; \beta)$
with $\beta > 0$ is finite if $\bar{f}, \bar{g} \in L^{2} (\mathbb{R}^3)$
or -- since Fourier transformation is an isometric isomorphism of $L^{2}
(\mathbb{R}^3)$ -- if $f, g \in L^{2} (\mathbb{R}^3)$.

If we perform the limit $\beta \to 0$ in the Fourier integral
(\ref{YukawaInt_f_g_FT}), we formally obtain:
\begin{equation}
  \label{CoulombInt_f_g_FT}
  \mathcal{C} (f, g) \; = \; 4\pi \, 
  \int \, \bigl[ \bar{f} (\bm{p}) \bigr]^{*} \, \frac{1}{p^{2}}
  \, \bar{g} (\bm{p}) \, \mathrm{d}^3 \bm{p} \, .
\end{equation}
The momentum space integral on the right-hand side can be interpreted to
be an inner product of the type of (\ref{Def_InnerProd}) that gives rise
to a suitable Hilbert space. 

The Cauchy-Schwarz Inequality (\ref{CauchyScwarz}) now requires $\bar{f}
(\bm{p})/p, \bar{g} (\bm{p})/p \in L^{2} (\mathbb{R}^3)$ to guarantee
that $\mathcal{C} (f, g)$ is finite. Obviously, this is much more
restrictive than the requirement $\bar{f}, \bar{g} \in L^{2}
(\mathbb{R}^3)$ or even $\bar{f} (\bm{p})/[\beta^{2}+p^{2}]^{1/2},
\bar{g} (\bm{p})/[\beta^{2}+p^{2}]^{1/2} \in L^{2} (\mathbb{R}^3)$, which
both guarantee that $\mathcal{Y} (f, g; \beta)$ is finite (alternative
criteria, which also guarantee the existence of the Coulomb integrals
$\mathcal{C} (f, g)$, are for instance formulated in \cite[Example 3
(Sobolev's inequality) on p.\ 31]{Reed/Simon/1975} or \cite[Section 4.3
(Hardy-Littlewood-Sobolev inequality)]{Lieb/Loss/1997}). 

This example clearly shows that the limit $\beta \to 0$ in integrals
involving the Yukawa potential $\exp (-\beta r)/r$ is not necessarily
continuous and does not always produce a finite result for arbitrary
square integrable charge densities $f$ and $g$.

If we set $\beta = 0$ in the Fourier transform (\ref{FT_YukPot}) of the
Yukawa potential, we formally obtain the Fourier transform of the Coulomb
potential (see for example \cite[Eq.\ (2) on p.\
194]{Gelfand/Vol1/Shilov/1964}):
\begin{equation}
  \label{FT_CoulombPot}
  (2\pi)^{-3/2} \int \, \frac{\exp (-\mathrm{i} \bm{p} \cdot
    \bm{r})}{r} \, \mathrm{d}^3 \bm{r} \; = \; 
  \frac{(2/\pi)^{1/2}}{p^{2}} \, .
\end{equation}
There is a fundamental difference between the Fourier transforms
(\ref{FT_YukPot}) and (\ref{FT_CoulombPot}). The Fourier integral in
(\ref{FT_YukPot}) is well defined and exists in the sense of classical
analysis, whereas the Fourier integral in (\ref{FT_CoulombPot}) diverges
in the sense of classical analysis and becomes mathematically meaningful
only if certain limiting or summation procedures are applied. Thus, the
Fourier transform (\ref{FT_YukPot}) of the Yukawa potential is a
\emph{function} in the ordinary sense.  In contrast, the Fourier
transform (\ref{FT_CoulombPot}) of the Coulomb potential is a
\emph{generalized function} or \emph{distribution} which is meaningful in
suitably restricted functionals only.

So, whenever we perform the limit $\beta \to 0$ in integrals involving
the Yukawa potential $\exp (-\beta r)/r$, we have to be cautious and take
into account that this limit may be discontinuous and that it does not
necessarily produce a finite result in the case of an essentially
arbitrary integrand. It will become clear in later Sections that these
problems are not restricted to integrals and that they can also occur in
orthogonal expansions of the Yukawa potential.

\typeout{==> Section: Divergent Orthogonal Expansions}
\section{Divergent Orthogonal Expansions}
\label{Sec:DivOrthExp}

Hilbert spaces, whose basic features are reviewed in Appendix
\ref{App:HilbertSpaces}, play a major role in various branches of
mathematics and mathematical physics and in particular also in
approximation theory. They also provide a rigorous mathematical framework
for quantum mechanics.

As discussed in Section \ref{Sec:OneRangAddThmCP} or in more details in
\cite[Section 3]{Weniger/2007b}, one-range addition theorems can also be
viewed to be special approximation procedures: A function $f (\bm{r} \pm
\bm{r}')$ belonging to suitable Hilbert space is expanded in terms of a
complete and orthonormal function set in such a way that the two argument
vectors $\bm{r}, \bm{r}' \in \mathbb{R}^{3}$ are separated. By
construction, such an orthogonal expansion converges in the mean with
respect to the norm of the corresponding Hilbert space.

Therefore, it certainly makes sense to discuss the basic properties of
orthogonal expansions in Hilbert spaces -- including their power as well
as their limitations -- in a relatively detailed way. Hilbert spaces are
linear vector spaces over the complex numbers equipped with an inner
product $( \cdot \vert \cdot )$ satisfying (\ref{Def_InnerProd}) and a
norm $\Vert \cdot \Vert$ satisfying (\ref{Def_Norm}), which has to be
\emph{finite}. A vector space with these properties is called a Hilbert
space if it is \emph{complete} with respect to its norm $\Vert \cdot
\Vert$.

Let us assume that $f$ is an element of some Hilbert space $\mathcal{H}$,
and that the functions $\{ \varphi_n \}_{n=0}^{\infty}$ are linearly
independent and complete in $\mathcal{H}$. Then, $f$ can be approximated
by \emph{finite} linear combinations
\begin{equation}
  \label{f_FinAppr}
  f_N \; = \; \sum_{n=0}^{N} C_{n}^{(N)} \varphi_n \, ,
  \qquad N \in \mathbb{N}_{0} \, .
\end{equation}
The coefficients $C_{n}^{(N)}$ are chosen in such a way that the mean
square deviation
\begin{equation}
  \label{Def_MeanSquareDeviation}
  \Vert f - f_N \Vert^2 \; = \; (f - f_N \vert f - f_N)  
\end{equation}
becomes minimal. 

The determination of the coefficients $C_{n}^{(N)}$ in (\ref{f_FinAppr})
by minimizing the mean square deviation (\ref{Def_MeanSquareDeviation})
only makes sense if both $f$ and the functions $\{ \varphi_{n}
\}_{n=0}^{\infty}$ are \emph{normalizable} according to $\Vert f \Vert <
\infty$ and $\Vert \varphi_{n} \Vert < \infty$, respectively. Thus, $f$
as well as the functions $\{ \varphi_{n} \}_{n=0}^{\infty}$ have to
belong to the Hilbert space $\mathcal{H}$.

The finite approximation (\ref{f_FinAppr}) converges to $f$ as $N \to
\infty$ if the mean square deviation (\ref{Def_MeanSquareDeviation}) can
be made as small as we like by increasing the summation limit $N$. In the
case of convergence, it looks natural to assume that $f$ possesses an
\emph{infinite expansion}
\begin{equation}
  \label{f_InfExp}
  f \; = \; \sum_{n=0}^{\infty} C_{n} \varphi_n
\end{equation}
in terms of the linearly independent and complete functions $\{ \varphi_n
\}_{n=0}^{\infty}$ with coefficients $C_{n} = \lim_{N \to \infty}
C_{n}^{(N)}$. 

Unfortunately, this is not true. In general, the coefficients
$C_{n}^{(N)}$ in (\ref{f_FinAppr}) do not only depend on $n$, $f$, and
$\{ \varphi_n \}_{n=0}^{\infty}$, but also on the summation limit $N$.
It is not \emph{a priori} clear whether the coefficients $C_{n}^{(N)}$ in
(\ref{f_FinAppr}) possess well defined limits $C_{n} = \lim_{N \to
  \infty} C_{n}^{(N)}$, or to put it differently, whether an infinite
expansion of the type of (\ref{f_InfExp}) exists. Expansions of the type
of (\ref{f_InfExp}) may or may not exist.

It is one of the central results of approximation theory that for
functions $f \in \mathcal{H}$ the mean square deviation
(\ref{Def_MeanSquareDeviation}) becomes minimal if the functions $\{
\varphi_n \}_{n=0}^{\infty}$ are not only linearly independent and
complete, but also \emph{orthonormal} satisfying $(\varphi_n \vert
\varphi_{n'}) = \delta_{n {n'}}$ for all indices $n, n' \in
\mathbb{N}_{0}$, and if the coefficients are chosen according to
$C_{n}^{(N)} = (\varphi_n \vert f)$ (see for example \cite[Theorem 9 on
p.\ 51]{Davis/1989}).

If the functions $\{ \varphi_n \}_{n=0}^{\infty}$ are complete and
orthonormal in $\mathcal{H}$ and if the expansion coefficients are chosen
according to $C_{n}^{(N)} = (\varphi_n \vert f)$, then the coefficients
$(\varphi_n \vert f)$ in $f_N$ do not depend on the truncation order $N$.
Thus, $f \in \mathcal{H}$ possesses an infinite series expansion
\begin{equation}
  \label{Expand_f_CONS}
f \; = \; \sum_{n=0}^{\infty} \, (\varphi_n \vert f) \, \varphi_n
\end{equation}
in terms of the complete and orthonormal function set $\{ \varphi_n
\}_{n=0}^{\infty}$, and this expansion converges in the mean with respect
to the norm $\Vert \cdot \Vert$ of the Hilbert space $\mathcal{H}$.

This is all well known and described in countless books on functional
analysis or approximation theory. In these books, it is always emphasized
that orthogonal expansions of the type of (\ref{Expand_f_CONS}) are
mathematically meaningful and converge in the mean if and only if $f \in
\mathcal{H}$.

However, in practical applications we are often confronted with functions
that are not normalizable and thus do not belong to the corresponding
Hilbert space $\mathcal{H}$. In some cases it may be desirable to expand
such a function $f \notin \mathcal{H}$ in terms of functions $\{
\varphi_n \}_{n=0}^{\infty}$, that are complete and orthonormal in
$\mathcal{H}$. It is thus a practically relevant question whether and
under which conditions the concept of orthogonal expansions in a Hilbert
space $\mathcal{H}$ can be extended to functions $f \notin \mathcal{H}$.

If this is indeed possible, we also have to analyze in which respect
orthogonal expansions of a function $f \in \mathcal{H}$ differ from those
of a function $f \notin \mathcal{H}$. In particular, we have to analyze
whether and under which conditions orthogonal expansions $f =
\sum_{n=0}^{\infty} (\varphi_n \vert f) \varphi_n$ can be used in inner
products $(f \vert g)$ with $f \notin \mathcal{H}$ and $g \in
\mathcal{H}$. 

There is one obvious complication: If $f \in \mathcal{H}$, then it
follows from the Cauchy-Schwarz Inequality (\ref{CauchyScwarz}) that the
map $g \mapsto (f \vert g)$ is bounded and thus continuous for all $g \in
\mathcal{H}$. If, however, $f \notin \mathcal{H}$, we have to take into
account the \emph{Riesz Representation Theorem} (see for example
\cite[Theorem 7.60 on p.\ 199]{Griffel/2002}):
\begin{quote}
  \sl For every continuous and thus bounded linear functional $U \colon
  \mathcal{H} \to \mathbb{C}$ there exists a unique $u \in \mathcal{H}$
  such that $U (v) = (u \vert v)$ for all $v \in \mathcal{H}$.
\end{quote}
Thus, the map $g \mapsto (f \vert g)$ with $f \notin \mathcal{H}$ cannot
be continuous and bounded for all $g \in \mathcal{H}$. Consequently, for
a given $f \notin \mathcal{H}$, there must be at least one $g \in
\mathcal{H}$ that yields an unbounded inner product $(f \vert g)$.

Accordingly, the following discussion has to be limited to those $g \in
\mathcal{H}$ that yield for a given $f \notin \mathcal{H}$ bounded inner
products $(f \vert g)$. Therefore, we have to assume that $g$ belongs to
the subset $\mathcal{F} \subset \mathcal{H}$ defined by
\begin{equation}
  \label{Def_F}
  \mathcal{F} \; = \; 
  \{ g \vert g \in \mathcal{H}, f \notin \mathcal{H}, \vert (f \vert g) 
  \vert < \infty \} \, .
\end{equation}
For arbitrary $f \notin \mathcal{H}$, it can happen that there is no $g
\in \mathcal{H}$ satisfying $\vert (f \vert g)\vert < \infty$, i.e., that
$\mathcal{F}$ is empty. In the following text, it will be assumed that
this is not the case. However, it follows from the Riesz Representation
Theorem that we cannot have $\mathcal{F} = \mathcal{H}$, i.e.,
$\mathcal{F}$ is either empty or a proper subset of $\mathcal{H}$.

A divergent orthogonal expansion $f = \sum_{n=0}^{\infty} (\varphi_n
\vert f) \varphi_n$ makes sense only if the inner products $(\varphi_n
\vert f)$ are finite for all finite indices $n$. If $f \in \mathcal{H}$,
this is guaranteed by the Cauchy-Schwarz Inequality (\ref{CauchyScwarz}),
but for $f \notin \mathcal{H}$ we have to assume explicitly that
$\varphi_{n} \in \mathcal{F}$ holds for all finite values of the index
$n$.

On the basis of these assumptions, it is at least formally possible to
construct an orthogonal expansion $f = \sum_{n=0}^{\infty} (\varphi_n
\vert f) \varphi_n$ even if $f \notin \mathcal{H}$. It is, however, not
at all clear whether and in which sense the formal series expansion
$\sum_{n=0}^{\infty} (\varphi_n \vert f) \varphi_n$ represents $f \notin
\mathcal{H}$. In particular, we have no \emph{a priori} reason to assume
that this expansion might converge in the mean according to the norm
$\Vert \cdot \Vert$ of $\mathcal{H}$, which would imply that
$\sum_{n=0}^{\infty} \vert (\varphi_n \vert f) \vert^{2} < \infty$ holds.

The divergence of the series $\sum_{n=0}^{\infty} \vert (\varphi_n \vert
f) \vert^{2}$ can be made plausible by analyzing the mean square
deviation of the difference between a function $f$ and its (possibly
divergent) orthogonal expansion $\sum_{n=0}^{\infty} (\varphi_{n} \vert
f) \varphi_{n}$: {\allowdisplaybreaks \begin{align}
  \label{MSD_f_CONS}
  & \Vert f - \sum_{n=0}^{\infty} \, (\varphi_{n} \vert f) \, \varphi_{n}
  \Vert^{2} \; = \; \Bigl( f - \sum_{n=0}^{\infty} \, (\varphi_{n} \vert
  f) \, \varphi_{n} \Bigm\vert f - \sum_{m=0}^{\infty} \, (\varphi_{m}
  \vert f) \, \varphi_{n} \Bigr)
  \notag \\
  & \qquad \; = \; (f \vert f) - \sum_{n=0}^{\infty} \, (\varphi_{n}
  \vert f)^{*} \, (\varphi_{n} \vert f) \notag
  \\
  & \qquad \qquad - \sum_{m=0}^{\infty} \, \sum_{n=0}^{\infty} \,
  (\varphi_{n} \vert f)^{*} \, (\varphi_{m} \vert f) \, (\varphi_{m}
  \vert \varphi_{n}) + \sum_{m=0}^{\infty} \, (\varphi_{m} \vert f) \, (f
  \vert \varphi_{m})
  \notag \\
  & \qquad \; = \; (f \vert f) - \sum_{m=0}^{\infty} \, \vert
  (\varphi_{m} \vert f) \vert^{2} \, .
\end{align}}%
If $f \in \mathcal{H}$, we have $\Vert f \Vert^{2} = (f \vert f) <
\infty$, and it follows from \emph{Parseval's Equality} (see for example
\cite[Eq.\ (II.2) on p.\ 45]{Reed/Simon/1980})
\begin{equation}
  \label{ParsevalEquality}
\Vert f \Vert^2 \; = \;
\sum_{n=0}^{\infty} \, \vert (\varphi_n \vert f) \vert^2
\end{equation}
that (\ref{MSD_f_CONS}) vanishes.

If, however, $f \notin \mathcal{H}$, then $\Vert f \Vert^{2} = (f \vert
f)$ diverges. In this case, the mean square deviation (\ref{MSD_f_CONS})
can only vanish if $\sum_{m=0}^{\infty} \vert (\varphi_{m} \vert f)
\vert^{2}$ also diverges. Of course, this is hand-waving and not a
rigorous mathematical proof. In particular, it is not at all clear
whether the mean square deviation (\ref{MSD_f_CONS}) makes sense at all
if $f \notin \mathcal{H}$. Nevertheless, this non-rigorous argument
should suffice to convince even a skeptical reader that it would be
overly optimistic to expect that the formal expansion
$\sum_{n=0}^{\infty} (\varphi_{n} \vert f) \varphi_{n}$ converges in the
mean to something finite if $f \notin \mathcal{H}$.

We should also take into account that series expansions -- or actually
all approximation schemes -- are practically useful only if they
reproduce the essential features of the function they represent. At least
in quantum mechanical bound state calculations, the norm of a function is
of considerable importance and should be preserved by an orthogonal
expansion. Otherwise, a function $f \notin \mathcal{H}$ could be
transformed to a function belonging to $\mathcal{H}$ simply be expanding
$f$ in terms of a complete and orthonormal function set. This would truly
be a miraculous achievement with revolutionary and most likely highly
undesirable consequences.

Plausibility arguments are no substitute for a rigorous proof. Such a
proof can be formulated with the help of the \emph{Riesz-Fischer Theorem}
(see for example \cite[Theorem 7.43 on p.\ 191]{Griffel/2002}):
\begin{quote}
  \sl Let $\{ \varphi_n \}_{n=0}^{\infty}$ be a complete and orthonormal
  function set in a Hilbert space $\mathcal{H}$, and let $c_{0}, c_{1},
  c_{2}, \dots$ be a sequence of numbers such that $\sum_{n=0}^{\infty}
  \vert c_{n} \vert^{2}$ converges. Then, the expansion
  $\sum_{n=0}^{\infty} c_{n} \varphi_{n}$ converges in the mean to some
  $w \in \mathcal{H}$ such that $c_{n} = (\varphi_{n} \vert w)$.
\end{quote}
This theorem shows that there is no sequence $(\varphi_{0} \vert f)$,
$(\varphi_{1} \vert f)$, $(\varphi_{2} \vert f)$, $\dots$ of inner
products such that $\Vert \sum_{n=0}^{\infty} (\varphi_{n} \vert f)
\varphi_{n} \Vert^{2} = \sum_{n=0}^{\infty} \vert (\varphi_{n} \vert f)
\vert^{2} < \infty$ and $f \notin \mathcal{H}$ simultaneously hold.
Accordingly, $f \notin \mathcal{H}$ implies that $\sum_{n=0}^{\infty}
(\varphi_{n} \vert f) \varphi_{n}$ diverges in the mean with respect to
the norm of $\mathcal{H}$. Thus, normalization is preserved by orthogonal
expansions even if $f \notin \mathcal{H}$.

This applies also to expansions in terms of Guseinov's functions
$\prescript{}{k}{\Psi}_{n, \ell}^{m} (\beta, \bm{r})$ defined by
(\ref{Def_Psi_Guseinov}). As discussed in \cite[p.\ 20]{Weniger/2007b},
the Yukawa potential belongs for $k = 0, 1, 2, \dots$ to the weighted
Hilbert space $L_{r^k}^{2} (\mathbb{R}^3)$ defined by
(\ref{HilbertL_r^k^2}), but not for $k=-1$ (see also \cite[p.\
410]{Homeier/Weniger/Steinborn/1992a}). Consequently, the expansion of
the Yukawa potential in terms of Guseinov's functions converges in the
mean for $k = 0, 1, 2, \dots$ and diverges for $k=-1$. In Appendix
\ref{App:ExpYukawa2GusFun}, it is explicitly shown that this is indeed
the case.

We thus arrive at the conclusion that \emph{conventional} Hilbert space
theory and the related concept of \emph{convergent} orthogonal expansions
of the type of (\ref{Expand_f_CONS}) in terms of complete and orthonormal
functions $\{ \varphi_n \}_{n=0}^{\infty}$ only make sense if $f \in
\mathcal{H}$.  Divergent expansions $f = \sum_{n=0}^{\infty}
(\varphi_{n}\vert f) \varphi_{n}$ with $f \notin \mathcal{H}$ are
\emph{generalized functions} in the sense of Schwartz
\cite{Schwartz/1966a} that can converge \emph{weakly} when used in
suitably restricted functionals.

In the context of one-range addition theorems, which are to be used in
multicenter integrals, it is essential that orthogonal expansions $f =
\sum_{n=0}^{\infty} (\varphi_{n} \vert f) \varphi_{n}$ can safely be used
in inner products $(f \vert g)$. If $f, g \in \mathcal{H}$, this can be
shown by applying the Cauchy-Schwarz Inequality (\ref{CauchyScwarz}) to
the inner product $\bigl( f - \sum_{n=0}^{N} (\varphi_{n} \vert f)
\varphi_{n} \bigm\vert g \bigr)$. We obtain:
\begin{equation}
  \label{Conv_InProd_CONS}
  \bigl\vert \bigl( f - \sum_{n=0}^{N} (\varphi_{n} \vert f) \varphi_{n} 
  \bigm\vert g \bigr)  \bigr\vert^{2} \; \leq \;  
  \Vert f - \sum_{n=0}^{N} (\varphi_{n} \vert f) \varphi_{n} \Vert^{2} \,
  \Vert g \Vert^{2} \, .
\end{equation}
The assumption $g \in \mathcal{H}$ implies $\Vert g \Vert^{2} < \infty$,
and the assumption $f \in \mathcal{H}$ implies that $\Vert f -
\sum_{n=0}^{N} (\varphi_{n} \vert f) \varphi_{n} \Vert^{2}$ vanishes as
$N \to \infty$. Accordingly, the right-hand side of
(\ref{Conv_InProd_CONS}) vanishes as $N \to \infty$.

If $g \in \mathcal{H}$ but $f \notin \mathcal{H}$, we cannot use the
Cauchy-Schwarz Inequality (\ref{CauchyScwarz}). Moreover, the inner
products $(f \vert g)$ is not necessarily finite. As discussed above, the
map $g \mapsto (f \vert g)$ with $f \notin \mathcal{H}$ cannot be
continuous and bounded for all $g \in \mathcal{H}$. Thus, it is essential
to assume that $g$ belongs to the subset $\mathcal{F} \subset
\mathcal{H}$ defined by (\ref{Def_F}).

Next, we have to formulate criteria, which guarantee that the divergent
orthogonal expansion $f = \sum_{n=0}^{\infty} (\varphi_{n} \vert f)
\varphi_{n}$ can safely be used in inner products $(f \vert g)$ with $g
\in \mathcal{F}$. Thus, we have to analyze under which conditions
\begin{equation}
  \label{f-f_CONS}
  \bigl( f - \sum_{n=0}^{\infty} (\varphi_{n} \vert f) \varphi_{n} 
  \bigm\vert g \bigr) \; = \; 0
\end{equation}
with $f \notin \mathcal{H}$ holds for suitable $g \in \mathcal{F}$.

There is a very simple situation in which (\ref{f-f_CONS}) is obviously
valid. Let us assume that there is some $\tilde{g} \in \mathcal{F}$ that
possesses a \emph{finite} expansion in terms of the complete and
orthonormal functions $\{ \varphi_n \}_{n=0}^{\infty}$:
\begin{equation}
  \label{Def_g_tilde}
  \tilde{g} \; = \; 
  \sum_{m=0}^{M} \, \tilde{\gamma}_{m} \, \varphi_{m} \; = \;
  \sum_{m=0}^{M} \, (\tilde{g} \vert \varphi_{m}) \, \varphi_{m} \, ,
  \qquad M \in \mathbb{N}_{0} \, .
\end{equation}
Since we always assume that the inner products $(\varphi_{n} \vert f)$
are finite for all finite values of $n$, we obtain:
\begin{align}
  & \bigl(f-\sum_{n=0}^{\infty} (\varphi_{n}\vert f) \varphi_{n}
  \bigm\vert \tilde{g}\bigr) \; = \; \bigl(f \bigm\vert \tilde{g}\bigr)
  \, - \, \Bigl( \sum_{n=0}^{\infty} (\varphi_{n} \vert f) \varphi_{n}
  \Bigm\vert \tilde{g} \Bigr)
  \notag \\
  & \qquad \; = \;
  \sum_{m=0}^{M} \, \tilde{\gamma}_{m} \, (f \vert \varphi_{m}) \, - \,
  \sum_{n=0}^{\infty} \, \sum_{m=0}^{M} \, (\varphi_{n}\vert f)^{*} \, 
  \tilde{\gamma}_{m} \, (\varphi_{n} \vert \varphi_{m})
  \notag \\
  & \qquad \; = \;
  \sum_{m=0}^{M} \, \tilde{\gamma}_{m} \, (f \vert \varphi_{m}) \, - \,
  \sum_{m=0}^{M} \, (\varphi_{m}\vert f)^{*} \, \tilde{\gamma}_{m}
  \; = \; 0 \, .
\end{align}
Thus, for functions $\tilde{g} \in \mathcal{F}$ satisfying
(\ref{Def_g_tilde}), the divergent orthogonal expansion $f =
\sum_{n=0}^{\infty} (\varphi_{n} \vert f) \varphi_{n}$ produces the
correct result in the inner product $(f \vert \tilde{g})$.

As a mild generalization of (\ref{Def_g_tilde}), let us now consider some
$g \in \mathcal{F}$ that possesses an \emph{infinite} expansion in terms
of the complete and orthonormal function $\{ \varphi_n
\}_{n=0}^{\infty}$:
\begin{equation}
  \label{Def_g_CONS}
  g \; = \; \sum_{m=0}^{\infty} \, \gamma_{m} \, \varphi_{m} \; = \;
  \sum_{m=0}^{\infty} \, (\varphi_{m} \vert g) \, \varphi_{m} \, .
\end{equation}
Since $g \in \mathcal{F} \subset \mathcal{H}$, this expansion converges
in the mean. However, the convergence of this expansion alone does not
suffice to guarantee that (\ref{f-f_CONS}) is satisfied. The problem is
that we are now confronted with infinite series that do not necessarily
converge:
\begin{align}
  \label{Gen_Conv_Cond_f}
  & \bigl(f-\sum_{n=0}^{\infty} (\varphi_{n}\vert f) \varphi_{n}
  \bigm\vert g \bigr) \; = \; \bigl(f \bigm\vert g \bigr)
  \, - \, \Bigl( \sum_{n=0}^{\infty} (\varphi_{n} \vert f) \varphi_{n}
  \Bigm\vert g \Bigr)
  \notag \\
  & \qquad \; = \; 
  \sum_{m=0}^{\infty} \, \gamma_{m} \, (f \vert \varphi_{m}) \, - \, 
  \sum_{n=0}^{\infty} \, (\varphi_{n}\vert f)^{*} \, 
  (\varphi_{n} \vert g)
  \notag \\
  & \qquad \; = \; 
  \sum_{m=0}^{\infty} \, \gamma_{m} \, (f \vert \varphi_{m}) \, - \, 
  \sum_{n=0}^{\infty} \, \sum_{m=0}^{\infty} \, (\varphi_{n}\vert f)^{*}
  \, \gamma_{m} \, (\varphi_{n} \vert \varphi_{m})
  \notag \\
  & \qquad \; = \; 
  \sum_{m=0}^{\infty} \, \gamma_{m} \, (f \vert \varphi_{m}) \, - \, 
  \sum_{n=0}^{\infty} \, (\varphi_{n}\vert f)^{*} \, \gamma_{n} \, .
\end{align}
If $\sum_{m=0}^{\infty} \gamma_{m} (f \vert \varphi_{m})$ and
$\sum_{n=0}^{\infty} (\varphi_{n}\vert f)^{*} \gamma_{n}$ both converge
to $(f \vert g)$, (\ref{f-f_CONS}) is satisfied, and the use of the
divergent orthogonal expansion $f = \sum_{n=0}^{\infty} (\varphi_{n}
\vert f) \varphi_{n}$ in the inner product $(f \vert g)$ produces the
correct result. 

The requirement, that the infinite series in (\ref{Gen_Conv_Cond_f}) have
to converge, makes it possible to characterize the subset $\mathcal{F}
\subset \mathcal{H}$ defined by (\ref{Def_F}) more precisely. The
expansion $g = \sum_{n=0}^{\infty} (\varphi_{n}\vert g) \varphi_{n}$
converges in the mean if the coefficients $(\varphi_{n}\vert g)$ decay
more rapidly than $n^{-1/2}$ as $n \to \infty$. Since $f \notin
\mathcal{H}$, the coefficients $(\varphi_{n}\vert f)$ of the expansion $f
= \sum_{n=0}^{\infty} (\varphi_{n}\vert f) \varphi_{n}$ either decay less
rapidly than $n^{-1/2}$ as $n \to \infty$ or they may even diverge as $n
\to \infty$. Thus, the coefficients $(\varphi_{n}\vert g)$ in
(\ref{Def_g_CONS}) have to decay so fast as $n \to \infty$ that the
infinite series $(f \vert g) = \sum_{n=0}^{\infty} (f \vert \varphi_{n})
(\varphi_{n} \vert g)$ converges. This is certainly the case if the
coefficients $(f \vert \varphi_{n}) (\varphi_{n} \vert g)$ decay more
rapidly than $1/n$ as $n \to \infty$.

Divergent orthogonal expansions $f = \sum_{n=0}^{\infty}
(\varphi_{n}\vert f) \varphi_{n}$ with $f \notin \mathcal{H}$ possess the
characteristic features of generalized functions in the sense of Schwartz
\cite{Schwartz/1966a}: Although divergent in the mean, such an expansion
is meaningful in inner products $(f \vert g)$ as long as $g$ is
restricted to the proper subset $\mathcal{F} \subset \mathcal{H}$ defined
by (\ref{Def_F}). If $g \notin \mathcal{F}$, the divergent orthogonal
expansion $f = \sum_{n=0}^{\infty} (\varphi_{n}\vert f) \varphi_{n}$
cannot be used in the inner product $(f \vert g)$ since the infinite
series $(f \vert g) = \sum_{m=0}^{\infty} (f \vert \varphi_{m})
(\varphi_{m} \vert g)$ diverges. 

Conceptually, this situation very much resembles the theory of rigged
Hilbert spaces or Gelfand triplets $\Phi \subset \mathcal{H} \subset
\Phi^{\times}$. Here, $\mathcal{H}$ is a Hilbert space, $\Phi$ is a
suitably restricted subset of $\mathcal{H}$, and $\Phi^{\times}$ is its
dual space defined by the condition that inner product $(u \vert v)$ with
$u \in \Phi^{\times}$ and $v \in \Phi$ remains finite. Loosely speaking,
we may say that the more we restrict the subset $\Phi \subset
\mathcal{H}$, the larger its dual space $\Phi^{\times}$ becomes. A very
readable account of rigged Hilbert spaces from the perspective of quantum
mechanics and their relationship with Dirac's bra and ket formalism can
be found in the book by Ballentine \cite[Chapter 1.4]{Ballentine/1998}.

The insight, that divergent orthogonal expansions $f =
\sum_{n=0}^{\infty} (\varphi_{n}\vert f) \varphi_{n}$ with $f \notin
\mathcal{H}$ are essentially generalized functions and can be used in a
mathematical rigorous way in inner products $(f \vert g)$ with $g \in
\mathcal{F}$, does not imply that all problems with the use of these
divergent series are solved. The characterization of the subset
$\mathcal{F} \subset \mathcal{H}$ is the crucial step that makes these
divergent expansions mathematically meaningful in inner products $(f
\vert g)$. But in real life applications, the characterization of
$\mathcal{F}$ may turn out be the most difficult problem that can occur
in this context.

\typeout{==> Section: One-Range Addition Theorems for the Coulomb
  Potential}
\section{One-Range Addition Theorems for the Coulomb Potential}
\label{Sec:OneRangAddThmCP}

Let us assume that $f$ belongs to the Hilbert space $L^2 (\mathbb{R}^3)$
of square integrable functions defined by (\ref{HilbertL^2}) and that the
functions $\{ \varphi_{n, \ell}^{m} (\bm{r}) \}_{n, \ell, m}$ are
complete and orthonormal in $L^2 (\mathbb{R}^3)$. As discussed in more
details in \cite[Section 3]{Weniger/2007b}, a one-range addition theorem
for $f (\bm{r} \pm \bm{r}')$, which converges in the mean with respect to
the norm of $L^2 (\mathbb{R}^3)$, can be constructed by expanding $f$ in
terms of the orthonormal functions $\{ \varphi_{n, \ell}^{m} (\bm{r})
\}_{n, \ell, m}$:
\begin{subequations}
  \label{OneRangeAddTheor}
  \begin{align}
     \label{OneRangeAddTheor_a}
    f (\bm{r} \pm \bm{r}') & \; = \; \sum_{n \ell m} \, C_{n, \ell}^{m}
    (f; \pm \bm{r}') \, \varphi_{n, \ell}^{m} (\bm{r}) \, ,
    \\
     \label{OneRangeAddTheor_b}
    C_{n, \ell}^{m} (f; \pm \bm{r}') & \; = \; \int \, \bigl[
    \varphi_{n, \ell}^{m} (\bm{r}) \bigr]^{*} \, f (\bm{r} \pm \bm{r}')
    \, \mathrm{d}^3 \bm{r} \, .
  \end{align}
\end{subequations}
The expansion (\ref{OneRangeAddTheor}) is indeed a one-range addition
theorem, since the variables ${\bm{r}}$ and ${\bm{r}}'$ are completely
separated: The dependence on $\bm{r}$ is entirely contained in the
functions $\varphi_{n, \ell}^{m} (\bm{r})$, whereas $\bm{r}'$ occurs only
in the expansion coefficients $C_{n, \ell}^{m} (f; \pm \bm{r}')$ which
are overlap integrals.

If the overlap integrals $C_{n, \ell}^{m} (f; \pm \bm{r}')$ can be
expanded in terms of the functions $\varphi_{n, \ell}^{m} (\bm{r}')$
according to
\begin{subequations}
  \label{OverlapExpand}
  \begin{align}
    \label{OverlapExpand_a}
    C_{n, \ell}^{m} (f; \pm \bm{r}') & \; = \; \sum_{n' \ell' m'} \,
    T_{n' \ell' m'}^{n \ell m} (f; \pm) \, \varphi_{n', \ell'}^{m'}
    (\bm{r}') \, ,
    \\
    \label{OverlapExpand_b}
    T_{n' \ell' m'}^{n \ell m} (f; \pm) & \; = \; \int \, \bigl[
    \varphi_{n', \ell'}^{m'} (\bm{r}') \bigr]^{*} \, C_{n, \ell}^{m} (f;
    \pm \bm{r}') \, \mathrm{d}^3 \bm{r}' \, ,
  \end{align}
\end{subequations}
then the addition theorem (\ref{OneRangeAddTheor}) assumes a completely
symmetrical form:
\begin{equation}
  \label{SymOneRangeAddTheor}
f (\bm{r} \pm \bm{r}') \; = \;
\sum_{\substack{n \ell m \\ n' \ell' m'}} \,
T_{n' \ell' m'}^{n \ell m} (f; \pm) \, \varphi_{n, \ell}^{m} (\bm{r}) \,
\varphi_{n', \ell'}^{m'} (\bm{r}') \, .
\end{equation}

As is well known in approximation theory, a nontrivial weight function $w
(\bm{r}) \ne 1$ can give more weight to those regions of space in which
$f$ is large, while deemphasizing the contribution from those regions in
which $f$ is small. Accordingly, the inclusion of a suitable weight
function $w \colon \mathbb{R}^{3} \to \mathbb{R}_{+}$ can improve
convergence. It is thus an obvious idea to construct one-range addition
theorems that converge with respect to the norm of a weighted Hilbert
space $L_{w}^{2} (\mathbb{R}^3)$ defined in (\ref{HilbertL_w^2}).  If $f
\in L_{w}^{2} (\mathbb{R}^3)$, we can construct a one-range addition
theorem by expanding $f (\bm{r} \pm \bm{r}')$ with respect to a function
set $\{ \psi_{n, \ell}^{m} (\bm{r}) \}_{n, \ell, m}$ that is complete in
$L_{w}^{2} (\mathbb{R}^3)$ and orthonormal with respect to the modified
inner product (\ref{InnerProd_w}) \cite[Eqs.\ (3.5) -
(3.7)))]{Weniger/2007b}.

In the articles \cite{Guseinov/2002c,%
  Guseinov/2002d,Guseinov/2003b,Guseinov/2003c,Guseinov/2003d,%
  Guseinov/2003e,Guseinov/2004a,Guseinov/2004b,Guseinov/2004c,%
  Guseinov/2004d,Guseinov/2004e,Guseinov/2004f,Guseinov/2004i,%
  Guseinov/2004k,Guseinov/2005a,Guseinov/2005b,Guseinov/2005c,%
  Guseinov/2005d,Guseinov/2005e,Guseinov/2005f,Guseinov/2005g,
  Guseinov/2006a,Guseinov/2006b,Guseinov/Aydin/Mamedov/2003,%
  Guseinov/Mamedov/2002d,Guseinov/Mamedov/2003,Guseinov/Mamedov/2004b,%
  Guseinov/Mamedov/2004d,Guseinov/Mamedov/2004e,%
  Guseinov/Mamedov/2004h,Guseinov/Mamedov/2005c,Guseinov/Mamedov/2005d,%
  Guseinov/Mamedov/2005g}, Guseinov and coworkers derived and applied
one-range addition theorems in connection with the special weight
function $w (r) = r^{k}$ with $k=-1, 0, 1, 2, \dots$ and used Guseinov's
functions $\prescript{}{k}{\Psi}_{n, \ell}^{m} (\beta, \bm{r})$ defined
by (\ref{Def_Psi_Guseinov}) as expansion functions. This yields one-range
addition theorems of the following general kind:
{\allowdisplaybreaks\begin{subequations}
  \label{SymOneRangeAddTheor_GusFun}
  \begin{align}
    \label{SymOneRangeAddTheor_GusFun_a}
    f (\bm{r} \pm \bm{r}') & \; = \; \sum_{\substack{n \ell m \\ n'
        \ell' m'}} \, \prescript{}{k}{\mathbf{T}}_{n' \ell' m'}^{n \ell
      m} (f; \beta, \pm) \,
    \prescript{}{k}{\Psi}_{n, \ell}^{m} (\beta, \bm{r}) \,
    \prescript{}{k}{\Psi}_{n', \ell'}^{m'} (\beta, \bm{r}') \, ,
    \\
    \label{SymOneRangeAddTheor_GusFun_b}
    \prescript{}{k}{\mathbf{T}}_{n' \ell' m'}^{n \ell m} (f; \beta,
    \pm) & \; = \; \int \, \bigl[ \prescript{}{k}{\Psi}_{n',
      \ell'}^{m'} (\beta, \bm{r}') \bigr]^{*} \, (r')^k \,
    \prescript{}{k}{\mathbf{C}}_{n, \ell}^{m} (f; \beta, \pm \bm{r}')
    \, \mathrm{d}^3 \bm{r}' \, ,
    \\
    \label{SymOneRangeAddTheor_GusFun_c}
    \prescript{}{k}{\mathbf{C}}_{n, \ell}^{m} (f; \beta, \pm \bm{r}')
    & \; = \; \int \, \bigl[ \prescript{}{k}{\Psi}_{n, \ell}^{m}
    (\beta, \bm{r}) \bigr]^{*} \, r^k \, f (\bm{r} \pm \bm{r}') \,
    \mathrm{d}^3 \bm{r} \, .
  \end{align}
\end{subequations}}
If $f \in L_{r^k}^{2} (\mathbb{R}^3)$, this addition theorem converges in
the mean according to the norm (\ref{InnerProd_r^k}) of the weighted
Hilbert space $L_{r^k}^{2} (\mathbb{R}^3)$.

As discussed in more details in \cite[Section 4]{Weniger/2007b}, Guseinov
derived in this way one-range addition theorems, for example for
Slater-type functions with integral and nonintegral principal quantum
numbers defined by (\ref{Def_STF}). For fixed $k = -1, 0, 1, 2, \dots$,
Guseinov's functions satisfy the orthogonality condition
(\ref{Psi_Guseinov_OrthoNor}) and they are complete and orthonormal in
the weighted Hilbert space $L_{r^k}^{2} (\mathbb{R}^3)$ defined by
(\ref{HilbertL_r^k^2}). As long as the Slater-type functions, which are
to be expanded, belong to the weighted Hilbert space $L_{r^k}^{2}
(\mathbb{R}^3)$, this is a completely legitimate approach that leads to
one-range addition theorems for Slater-type functions which converge in
the mean with respect to the norm (\ref{Norm_r^k_2}) of $L_{r^k}^{2}
(\mathbb{R}^3)$.

The Coulomb potential plays a central role in electronic structure
calculations, and the evaluation of inter-electronic repulsion integrals
of the type of (\ref{CouInt_f_g}) leads to formidable computational
problems, in particular if the densities $f$ and $g$ in
(\ref{CouInt_f_g}) are two-center charge densities of the type of $u
(\bm{r}-\bm{A}) v (\bm{r}-\bm{B})$ with $\bm{A}, \bm{B} \in
\mathbb{R}^{3}$. Therefore, it would be desirable to have a one-range
addition theorem for the Coulomb potential. In \cite{Guseinov/2005a},
Guseinov tried to accomplish this by expanding $1/ \vert \bm{r} - \bm{r}'
\vert$ in terms of functions $\prescript{}{k}{\Psi}_{n, \ell}^{m} (\beta,
\bm{r})$. Formally, Guseinov's approach leads to the following
symmetrical one-range addition theorems: 
{\allowdisplaybreaks
  \begin{subequations}
    \label{CP_OnerangeAddThm_GusFun_k}
    \begin{align}
      \label{CP_OnerangeAddThm_GusFun_k_a}
      \frac{1}{\vert \bm{r} - \bm{r}' \vert} & \; = \; \sum_{\substack{n
          \ell m \\ n' \ell' m'}} \, \prescript{}{k}{\mathbf{\Gamma}}_{n'
        \ell' m'}^{n \ell m} (\beta) \, \prescript{}{k}{\Psi}_{n,
        \ell}^{m} (\beta, \bm{r}) \, \prescript{}{k}{\Psi}_{n',
        \ell'}^{m'} (\beta, \bm{r}') \, ,
      \\
      \label{CP_OnerangeAddThm_GusFun_k_b}
      \prescript{}{k}{\mathbf{\Gamma}}_{n' \ell' m'}^{n \ell m} (\beta) &
      \; = \; \int \, \bigl[ \prescript{}{k}{\Psi}_{n', \ell'}^{m'}
      (\beta, \bm{r}') \bigr]^{*} \, r'^k \,
      \prescript{}{k}{\mathbf{C}}_{n, \ell}^{m} (\beta, \bm{r}') \,
      \mathrm{d}^3 \bm{r}' \, ,
      \\
      \label{CP_OnerangeAddThm_GusFun_k_c}
      \prescript{}{k}{\bm{\mathcal{C}}}_{n, \ell}^{m} (\beta, \bm{r}') &
      \; = \; \int \, \bigl[ \prescript{}{k}{\Psi}_{n, \ell}^{m} (\beta,
      \bm{r}) \bigr]^{*} \, \frac{r^k}{\vert \bm{r} - \bm{r}' \vert} \,
      \mathrm{d}^3 \bm{r} \, .
    \end{align}
  \end{subequations}}%
There are, however, some principal problem with these orthogonal
expansions which Guseinov had either overlooked or ignored in his earlier
work and which he still ignored in his most recent preprints
\cite{Guseinov/2007a,Guseinov/2007b,Guseinov/2007c}, although I had
emphasized their importance in \cite{Weniger/2007b}. In order to convince
both Guseinov as well as other skeptical readers, I presented in Section
\ref{Sec:DivOrthExp} a detailed discussion of the properties of
orthogonal expansions. The central features of these expansions can be
summarized as follows:
\begin{enumerate}
\item If $f$ belongs to some Hilbert space $\mathcal{H}$, then $f$ can be
  expanded in terms of a function set $\{ \varphi_n \}_{n=0}^{\infty}$
  that is complete and orthonormal in $\mathcal{H}$, and the expansion
  $\sum_{n=0}^{\infty} (\varphi_n \vert f) \varphi_n$ converges to $f$ in
  the mean with respect to the norm of $\mathcal{H}$.
\item If $f \notin \mathcal{H}$, the formal orthogonal expansion $f =
  \sum_{n=0}^{\infty} (\varphi_n \vert f) \varphi_n$ diverges in the mean
  with respect to the norm of $\mathcal{H}$.
\item Nevertheless, such a divergent orthogonal expansion $f =
  \sum_{n=0}^{\infty} (\varphi_n \vert f) \varphi_n$ can produce
  meaningful results in inner products $(f \vert g)$ as long as $g$ is
  restricted to the proper subset $\mathcal{F} \subset \mathcal{H}$
  defined by (\ref{Def_F}).
\end{enumerate}

The weighted Hilbert spaces $L_{r^k}^{2} (\mathbb{R}^3)$ with $k = -1, 0,
1, 2, \dots$ are based on the inner product (\ref{InnerProd_r^k}) which
involves an integration over the whole three-dimensional space
$\mathbb{R}^{3}$ with weight function $w (r) = r^{k}$. The Coulomb
potential $1/\vert \bm{r} - \bm{r}' \vert$ does not belong to any of the
weighted Hilbert spaces $L_{r^k}^{2} (\mathbb{R}^3)$ which Guseinov
implicitly used in his work. This implies that the one-range addition
theorems (\ref{CP_OnerangeAddThm_GusFun_k}) of the Coulomb potential
diverge for $k = -1, 0, 1, 2, \dots$ in the mean with respect to the
norms (\ref{Norm_r^k_2}) of the weighted Hilbert spaces $L_{r^k}^{2}
(\mathbb{R}^3)$.

Guseinov was not the first one who had derived a divergent expansion for
the Coulomb potential in terms of a complete and orthonormal function
set. Salmon, Birss, and Ruedenberg \cite{Salmon/Birss/Ruedenberg/1968}
derived a bipolar expansion of the Coulomb potential in terms of the
Gaussian-type eigenfunctions of a three-dimensional isotropic harmonic
oscillator which are complete and orthonormal in the Hilbert space $
L^{2} (\mathbb{R}^3)$ of square integrable functions. However,
Silverstone and Kay \cite{Silverstone/Kay/1969} demonstrated that this
expansion diverges, which was confirmed by Ruedenberg and Salmon
\cite{Ruedenberg/Salmon/1969}. Apparently, this observation was some kind
of death sentence for the bipolar expansion of Salmon, Birss, and
Ruedenberg \cite{Salmon/Birss/Ruedenberg/1968}. As far as I know, nobody
has ever used this expansion.

The divergence of the one-range addition theorems
(\ref{CP_OnerangeAddThm_GusFun_k}) for the Coulomb potential cannot be
ignored. So, we are confronted with the question what we should do with
these addition theorems. We could dismiss them as practically useless and
ignore them, as it was done with the divergent bipolar expansion of
Salmon, Birss, and Ruedenberg \cite{Salmon/Birss/Ruedenberg/1968}.
However, I think that this would be premature. Salmon, Birss, and
Ruedenberg published their divergent bipolar expansion in 1968, and we
now have a much better understanding of generalized functions and we also
know much more about numerically efficient summation techniques which
often can associate a finite value to a divergent series (see also the
discussion in \cite[pp.\ 24 - 25]{Weniger/2007b}).

As discussed in Section \ref{Sec:DivOrthExp}, divergent expansions of the
type of (\ref{CP_OnerangeAddThm_GusFun_k}) can nevertheless be
practically useful as long as they are exclusively used in suitably
restricted functionals. Thus, these functions are essentially generalized
functions in the sense of Schwartz \cite{Schwartz/1966a}. Obviously, this
offers new perspectives, but one must not forget that in relationships
involving generalized functions one always has to be extremely careful
about their domains of validity.

Consequently, it is not acceptable to proceed like Guseinov and to ignore
the distributional nature of divergent one-range addition theorems of the
type of (\ref{CP_OnerangeAddThm_GusFun_k}) and to treat them like
ordinary orthogonal expansions that converge in the mean. It is
absolutely essential to formulate regularity conditions for functionals
-- in our case multicenter integrals -- and to take them into account.
Otherwise, the use of distributional one-range addition theorems in
multicenter integrals would be purely experimental. Virtually every
outcome would be possible, depending on the other functions occurring in
the integral. The fact that Guseinov apparently encountered no problems
in his numerical examples does not prove anything.

I do not expect that it will be easy to formulate the necessary
regularity conditions. Firstly, the Coulomb potential is in spite of its
apparent simplicity a relatively complicated mathematical object (see for
example \cite[Chapter 9]{Lieb/Loss/1997}). Secondly, I fear that for
every type of multicenter integral containing the Coulomb potential a new
set of regularity conditions has to be formulated. 

In \cite[p.\ 23]{Weniger/2007b}), a possible strategy based on orthogonal
expansions of the charge densities $f$ and $g$ in Coulomb integrals
$\mathcal{C} (f, g)$ defined by (\ref{CouInt_f_g}) was sketched. It
cannot be denied that this approach would be highly pedestrian, and more
elegant and more powerful alternatives would be highly desirable. It
seems that a lot of work remains to be done before distributional
one-range addition theorems of the type of
(\ref{CP_OnerangeAddThm_GusFun_k}) can safely and effectively be applied
in multicenter integrals.

The idea of using distributional orthogonal expansions, which diverge in
the mean and converge only weakly in suitably restricted functionals, is
not new. In \cite{Weniger/1985}, I derived expansions of the plane wave
$\exp (\mathrm{i} \bm{p} \cdot \bm{r})$ in terms of complete orthonormal
and biorthogonal function sets that converge only weakly. In some cases,
these expansions simplify the evaluation of Fourier transforms, and they
can also be used for the construction of one-range addition theorems (see
\cite{Homeier/Weniger/Steinborn/1992a} or \cite[Section
VII]{Weniger/1985}).

Expansions for the plane wave, that closely resemble those derived in
\cite{Weniger/1985}, were also constructed by Guseinov \cite[Eqs.\ (45) -
(46)]{Guseinov/2003c}. Guseinov, who did not mention \cite{Weniger/1985}
in \cite{Guseinov/2003c}, either overlooked or deliberately ignored the
obvious fact that that the plane wave does not belong to any of the
Hilbert spaces which he implicitly used. Accordingly, Guseinov's
expansion diverge in the mean and can only converge weakly. Guseinov's
oversight is hard to understand because he had cited \cite{Weniger/1985}
in several other articles
\cite{Guseinov/2002c,Guseinov/2004c,Guseinov/2004d,Guseinov/2005b,%
  Guseinov/2005c,Guseinov/Mamedov/2001c,Guseinov/Mamedov/2002d,%
  Guseinov/Mamedov/2004e}.

\typeout{==> Section: Guseinov's Rearranged Addition Theorems}
\section{Guseinov's Rearranged Addition Theorems}
\label{Sec:RearrAddThm}

As discussed in Section \ref{Sec:DivOrthExp}, orthogonal expansions play
a central role in Hilbert spaces and also in approximation theory. In
contrast, nonorthogonal expansions are largely ignored. Of course, there
are many good reasons for this neglect. In the context of one-range
addition theorems, the most important evidence speaking against the use
of nonorthogonal function is the following well established fact: If a
function set $\{ \varphi_n \}_{n=0}^{\infty}$ is only complete in a given
Hilbert space $\mathcal{H}$, but not orthogonal, then it is general only
possible to construct finite approximations to $f \in \mathcal{H}$ of the
type of (\ref{f_FinAppr}) by minimizing the mean square deviation
(\ref{Def_MeanSquareDeviation}), but the existence of formal expansions
of the type of (\ref{f_InfExp}) in terms of nonorthogonal functions is
not guaranteed: Thus, nonorthogonal expansions may or may not exist. This
fact is documented quite extensively in the mathematical literature (see
for example \cite[Theorem 10 on p.\ 54]{Davis/1989} or \cite[Section
1.4]{Higgins/1977}) or also in the literature on electronic structure
calculations \cite{Klahn/1975,Klahn/Bingel/1977a,Klahn/Bingel/1977b,%
  Klahn/Bingel/1977c,Klahn/1981,Klahn/Morgan/1984}). Horrifying examples
of pathologies of nonorthogonal expansions can be found in \cite[Section
III.I]{Klahn/1981}.

Of course, there are situations in which nonorthogonal expansions offer
computational advantages (see fore example the discussion in
\cite{Daubechies/Grossmann/Meyer/1986}). However, in the vast majority of
all cases, orthogonal expansions have clearly superior properties.
Consequently, one should not voluntarily abandon the highly useful
feature of orthogonality unless there are truly compelling reasons.

In \cite{Guseinov/2001,Guseinov/2002c}, Guseinov derived one-range
addition theorems for Slater-type functions with integral and nonintegral
principal quantum numbers of the type of (\ref{OneRangeAddTheor}) by
expanding them in terms of his complete and orthonormal functions
$\prescript{}{k}{\Psi}_{n, \ell}^{m} (\beta, \bm{r})$. As long as the
Slater-type functions belong to the weighted Hilbert spaces $L_{r^k}^{2}
(\mathbb{R}^3)$ defined by (\ref{HilbertL_r^k^2}) with $k=-1, 0, 1, 2,
\dots$, these addition theorem converge in the mean with respect to the
norms (\ref{Norm_r^k_2}) of these Hilbert spaces.

For reasons, which I do not really understand, Guseinov considered it to
be advantageous to replace in his one-range addition theorems for
Slater-type functions his complete and orthonormal functions
$\prescript{}{k}{\Psi}_{n, \ell}^{m} (\beta, \bm{r})$ by nonorthogonal
Slater-type functions with integral principal quantum numbers via
\cite[Eq.\ (6.4)]{Weniger/2007b}
\begin{align}
  \label{GusFun2STF}
  & \prescript{}{k}{\Psi}_{n, \ell}^{m} (\beta,
  \bm{r}) \; = \; 2^{\ell} \, \left[ \frac{(2\beta)^{k+3} \,
      (n+\ell+k+1)!}{(n-\ell-1)!} \right]^{1/2}
  \notag \\
  & \qquad \times \, \sum_{\nu=0}^{n-\ell-1} \,
  \frac{(-n+\ell+1)_{\nu} \, 2^{\nu}}{(2\ell+k+\nu+2)! \, \nu!} \,
  \chi_{\nu+\ell+1, \ell}^{m} (\beta, \bm{r})
\end{align}
and to rearrange the order of summations of the resulting expansions.
Guseinov constructed in this way expansions of Slater-type functions
$\chi_{N, L}^{M} (\beta, \mathbf{r} \pm \mathbf{r}')$ with in general
nonintegral principal quantum numbers $N \in \mathbb{R} \setminus
\mathbb{N}$ in terms of Slater-type functions $\chi_{n, \ell}^{m} (\beta,
\mathbf{r})$ with integral principal $n \in \mathbb{N}$ quantum numbers
located at a different center (see also \cite[Section 6]{Weniger/2007b}).

As is well known, Slater-type functions are complete in all Hilbert space
implicitly used by Guseinov (for a proof, see \cite[Section
4]{Klahn/Bingel/1977b}), but not orthogonal. In view of the principal
problems mentioned above, it is therefore not at all clear whether
Guseinov's rearranged addition theorems are mathematically meaningful.
Accordingly, I claimed in \cite[Section 6]{Weniger/2007b}) that
Guseinov's rearrangements of his one-range addition theorems, which are
expansions in terms of his functions $\prescript{}{k}{\Psi}_{n, \ell}^{m}
(\beta, \bm{r})$ and thus ultimately expansions in terms of generalized
Laguerre polynomials $L_{n-\ell-1}^{(2\ell+k+2)} (2\beta r)$, are
dangerous and potentially disastrous and that their validity has to be
checked.

In \cite[p.\ 7]{Guseinov/2007a}, Guseinov disagreed and claimed this his
Eq.\ (3.11) -- a finite nested sum containing his functions
$\prescript{}{k}{\Psi}_{n, \ell}^{m} (\beta, \bm{r})$ on the left-hand
side and Slater-type functions $\chi_{n, \ell}^{m} (\beta, \bm{r})$ with
integral principal quantum numbers on the right-hand side -- proves the
validity and mathematical soundness of his approach.

I do not question the validity of Guseinov's Eq.\ (3.11), but I very much
disagree with Guseinov's conclusion that his Eq.\ (3.11) proves the
validity of his rearrangements. The problem with Guseinov's reasoning is
that he does not distinguish carefully between rearrangements of
\emph{finite} and \emph{infinite} sums. Obviously, a finite sum
\begin{equation}
  \label{FinSum_GLag}
  F_{N} (x) \; = \; \sum_{n=0}^{N} \, 
  \lambda_{n}^{(\alpha)} \, L_{n}^{(\alpha)} (x) \, ,
  \qquad N \in \mathbb{N}_{0} \, ,
\end{equation}
of generalized Laguerre polynomials $L_{n}^{(\alpha)} (x)$ multiplied by
purely numerical coefficients $\lambda_{n}^{(\alpha)}$ can always be
rearranged. If we insert the explicit expression (\ref{Def_GLagPol}) of
the generalized Laguerre polynomials into (\ref{FinSum_GLag}) and
rearrange the order of summations, we obtain
\begin{equation}
  \label{RearrFinSum_GLag}
  F_{N} (x) \; = \; \sum_{\nu=0}^{N} \, \frac{(-x)^{\nu}}{\nu!} \,
  \sum_{\mu=0}^{N-\nu} \, \frac{(\alpha+\nu+1)_{\mu}}{\mu!} \,  
  \lambda_{\mu+\nu}^{(\alpha)} \, .
\end{equation}
But if we now perform in the finite sum (\ref{FinSum_GLag}) the limit $N
\to \infty$ and consider instead the rearrangement of the infinite series
\begin{equation}
  \label{InfSum_GLag}
  F (x) \; = \; \lim_{N \to \infty} \, F_{N} (x) \; = \; 
  \sum_{n=0}^{\infty} \, 
  \lambda_{n}^{(\alpha)} \, L_{n}^{(\alpha)} (x) \, , 
\end{equation}
the situation is much more complicated and many things can go wrong.
Formally, a rearrangement of $F (x)$ yields the following power series in
$x$:
\begin{equation}
  \label{RearrInfSum_GLag}
  F (x) \; = \;  
  \sum_{\nu=0}^{\infty} \, \frac{(-x)^{\nu}}{\nu!} \,
  \sum_{\mu=0}^{\infty} \, \frac{(\alpha+\nu+1)_{\mu}}{\mu!} \,  
  \lambda_{\mu+\nu}^{(\alpha)} \, .  
\end{equation}
This power series for $F (x)$ makes sense if and only if the inner series
on the right-hand side of (\ref{RearrInfSum_GLag}) converges for every
$\nu \in \mathbb{N}_{0}$. In addition, the inner series in $\mu$ has to
produce values that do not increase too strongly with increasing $\nu$,
because otherwise the right-hand side of (\ref{RearrInfSum_GLag})
diverges for every $\vert x \vert > 0$. In the case of an essentially
arbitrary function $F (x)$, these two conditions are not necessarily
satisfied. It is also easy to show that the convergence of the Laguerre
expansion for $F (x)$ with respect to the norm of the Laguerre-type
Hilbert space $L^{2}_{\mathrm{e}^{-x} x^{\alpha}} (\mathbb{R}_{+})$
defined by (\ref{HilbertL^2_Lag}) does not imply the convergence of the
inner series over $\mu$ on the right-hand side of
(\ref{RearrInfSum_GLag}) for every $\nu \in \mathbb{N}_{0}$.

Special attention deserves the case that $F (x)$ is not an
\emph{analytic function} in the sense of complex analysis at the
expansion point $x=0$. In this case, the rearranged power series on
right-hand side of (\ref{RearrInfSum_GLag}) cannot exist because all but
a finite number of series coefficients are infinite, even if the Laguerre
expansion for $F (x)$ exists and converges in the mean with respect to
the norm of $L^{2}_{\mathrm{e}^{-x} x^{\alpha}} (\mathbb{R}_{+})$.

Accordingly, I see no reason to alter my assessment \cite[p.\
18]{Weniger/2007b} that Guseinov's rearrangements are dangerous and
potentially disastrous, and that their validity must be checked
explicitly.

The rearrangements of the finite sum $F_{N} (z)$ or of the infinite
series $F (z)$ yielding (\ref{RearrFinSum_GLag}) and
(\ref{RearrInfSum_GLag}), respectively, are special cases of the
rearrangements of double series $\sum_{m=0}^{\infty} \sum_{n=0}^{\infty}
a_{m, n}$. This is an old and extensively studied topic in the theory of
infinite series. The most detailed treatment, which I am aware of, can be
found in the book by Bromwich \cite[Chapter V]{Bromwich/1991}. Loosely
speaking, the rearrangement of such a double series is safe if the double
series converges absolutely. In the case of expansions in terms of
orthogonal polynomials, we cannot tacitly assume absolute converge.

So, we have good reason to assume that Guseinov's rearrangements of
one-range addition theorems are dangerous and potentially disastrous. Of
course, it is not satisfactory if we only know that a given mathematical
manipulation is dangerous. Instead, we would like to know with certainty
whether this operation is legitimate or not.

Unfortunately, one-range additions theorems for exponentially decaying
functions are fairly complicated mathematical objects, and
\emph{explicit} proofs of their convergence and/or divergence are very
difficult. Most likely, such an investigation would be a nontrivial
research problem in its own right. Since I am convinced that Guseinov's
rearrangements are basically a bad idea, I saw no point in spending too
much time and effort. Therefore, I looked for simpler alternatives to a
detailed convergence analysis, even if these alternatives would not
answer all questions of interest.

As shown in \cite[Section 6]{Weniger/2007b}, valuable insight can in some
cases be gained by considering not the complicated one-range addition
theorems themselves, but their much simpler one-center limits. Let us
therefore assume that we succeeded in constructing a one-range addition
theorem of the type of (\ref{SymOneRangeAddTheor_GusFun}) for some
function $f (\bm{r} \pm \bm{r}')$ by expanding it in terms of Guseinov's
functions.  If we now consider its one-center limit by setting $\bm{r}' =
\bm{0}$, our addition theorem must simplify to yield an expansion of $f
(\bm{r})$ in terms of Guseinov's functions:
\begin{subequations}
  \label{exp_f2gPsi}
  \begin{align}
    f (\bm{r}) & \; = \; \sum_{n \ell m} \,
    \prescript{}{k}{\mathcal{F}}_{n, \ell}^{m} (\beta; f) \,
    \prescript{}{k}{\Psi}_{n, \ell}^{m} (\beta, \bm{r}) \, ,
    \\
    \prescript{}{k}{\mathcal{F}}_{n, \ell}^{m} (\beta; f) & \; = \; \int
    \, \left[ \prescript{}{k}{\Psi}_{n, \ell}^{m} (\beta, \bm{r})
    \right]^{*} \, r^k \, f (\bm{r}) \, \mathrm{d} \bm{r} \, .
  \end{align}
\end{subequations}
Under fortunate circumstances, the mathematical nature of such an
identity allows conclusions about the legitimacy of Guseinov's
rearrangements.

Let us now assume that we succeeded in deriving a one-range addition
theorem of the type of (\ref{SymOneRangeAddTheor_GusFun}) by expanding
Slater-type functions $\chi_{N, L}^{M} (\beta, \mathbf{r} \pm \bm{r}')$
with in general nonintegral principal quantum numbers $N \in \mathbb{R}
\setminus \mathbb{N}$ in terms of Guseinov's functions
$\prescript{}{k}{\Psi}_{n, \ell}^{m} (\beta, \bm{r})$ with \emph{equal}
scaling parameters $\beta > 0$ (see for example \cite[Eq.\
(6.1)]{Weniger/2007b} with $\beta=\gamma$). If we now set $\bm{r}' =
\bm{0}$ in this addition theorem, it must simplify to yield the following
expansion of $\chi_{N, L}^{M} (\beta, \mathbf{r})$ in terms of Guseinov's
functions:
\begin{align}
  \label{NISTF2Gusfun_EqScaPar}
  & \chi_{N, L}^{M} (\beta, \bm{r}) \; = \; \frac
  {(2\gamma)^{-(k+3)/2}}{2^{N-1}} \, \Gamma (N+L+k+2)
  \notag \\
  & \qquad \times \, \sum_{\nu=0}^{\infty} \,
  \frac{(-N+L+1)_{\nu}}{\bigl[ (\nu+2L+k+2)! \, \nu! \bigr]^{1/2}}
  \, \prescript{}{k}{\Psi}_{\nu+L+1, L}^{M} (\beta, \bm{r}) \, ,
  \notag \\
  & \qquad \qquad N \in \mathbb{R} \setminus \mathbb{N} \, ,
    \qquad \beta > 0 \, , \qquad k = -1, 0, 1, 2, \dots \, .
\end{align}
If $N \in \mathbb{N}$ and $N \ge L+1$, the infinite series on the
right-hand side terminates because of the Pochhammer symbol
$(-N+L+1)_{\nu}$.

It is easy to show that the expansion (\ref{NISTF2Gusfun_EqScaPar}) in
terms of Guseinov's function is a special case of the following expansion
\cite[Eq.\ (5.17)]{Weniger/2007b} which expresses a nonintegral power
$x^{\mu}$ with $\mu \in \mathbb{R} \setminus \mathbb{N}_0$ as an infinite
series of generalized Laguerre polynomials:
\begin{align}
  \label{GenPow2GLag}
  x^{\mu} & \; = \; \frac{\Gamma (\mu+\alpha+1)}{\Gamma (\alpha+1)} \,
  \sum_{n=0}^{\infty} \, \frac{(-\mu)_n}{(\alpha+1)_n} \,
  L_{n}^{(\alpha)} (x) \, ,
  \notag \\
  & \qquad \mu \in \mathbb{R} \setminus \mathbb{N}_0 \, , \qquad \Re
  (\mu+\alpha) > - 1  \, , \qquad \Re (\alpha) > - 1 \, .
\end{align}
If we set $\mu=m$ with $m \in \mathbb{N}_0$, the infinite series on the
right-hand side terminates because of the Pochhammer symbol $(-m)_n$.

In \cite[Eqs.\ (6.9) - (6.11)]{Weniger/2007b}, it was shown that it is
not possible to transform the Laguerre expansion (\ref{GenPow2GLag}) for
$x^{\mu}$ with $\mu \in \mathbb{R} \setminus \mathbb{N}_0$ to a power
series in $x$ by inserting the explicit expression (\ref{Def_GLagPol}) of
the generalized Laguerre polynomials. Interchanging the order of the
nested summations yields a formal power series in $x$ \cite[Eq.\
(6.9)]{Weniger/2007b}. Superficially, this looks like success. However,
the coefficients of this power series can be expressed as hypergeometric
series ${}_{1} F_{0}$ which are for all but a finite number of indices
\emph{infinite} \cite[Eq.\ (6.11)]{Weniger/2007b}.

Of course, this failure is not really surprising: The general power
function $z^{\mu}$ with $z \in \mathbb{C}$ and $\mu \in \mathbb{C}
\setminus \mathbb{N}_0$ is not analytic at $z=0$ in the sense of complex
analysis. For $\mu=m$ with $m \in \mathbb{N}_0$, Taylor expansion of
$z^m$ around $z=0$ is, however, legitimate and yields the trivial
identity $z^m = z^m$.

Thus, we can conclude that in the case of equal scaling parameters $\beta
> 0$ the one-center limit $\bm{r}' = \bm{0}$ of Guseinov's rearranged
addition theorem for Slater-type functions $\chi_{N, L}^{M} (\beta,
\bm{r} \pm \bm{r}')$ does not exist if the principal quantum number $N$
is nonintegral, $N \in \mathbb{R} \setminus \mathbb{N}$.

Let us now assume that we succeeded in deriving a one-range addition
theorem of the type of (\ref{SymOneRangeAddTheor_GusFun}) by expanding
Slater-type functions $\chi_{N, L}^{M} (\beta, \mathbf{r} \pm \bm{r}')$
with in general nonintegral principal quantum numbers $N \in \mathbb{R}
\setminus \mathbb{N}$ in terms of Guseinov's functions
$\prescript{}{k}{\Psi}_{n, \ell}^{m} (\gamma, \bm{r})$ with different
scaling parameters $\beta \ne \gamma > 0$ (see for example \cite[Eq.\
(6.1)]{Weniger/2007b}). If we now set $\bm{r}' = \bm{0}$ in this addition
theorem, it must simplify to yield the following expansion of $\chi_{N,
  L}^{M} (\beta, \mathbf{r})$ in terms of Guseinov's functions with
different scaling parameter $\gamma \ne \beta > 0$:
\begin{align}
  \label{Expand_NISTF2Gusfun_DiffScaPar_1}
  & \chi_{N, L}^{M} (\beta, \bm{r}) \; = \; \frac
  {(2\gamma)^{L+(k+3)/2} \, \beta^{N-1}}{[\beta+\gamma]^{N+L+k+2}}
  \, \frac{\Gamma (N+L+k+2)}{(2L+k+2)!}
  \notag \\
  & \qquad \times \sum_{\nu=0}^{\infty} \, \left[
    \frac{(\nu+2L+k+2)!}{\nu!} \right]^{1/2} \,
  \prescript{}{k}{\Psi}_{\nu+L+1, L}^{M} (\gamma, \bm{r})
  \notag \\
  & \qquad \qquad \times {}_2 F_1 \left(-\nu, N+L+k+2; 2L+k+3;
    \frac{2\gamma}{\beta+\gamma} \right) \, ,
  \notag \\
  & \qquad \qquad \qquad N \in \mathbb{R} \setminus \mathbb{N} \, , 
  \qquad \beta, \gamma > 0 \, .
\end{align}
If we set $\gamma = \beta$, we of course obtain
(\ref{NISTF2Gusfun_EqScaPar}). 

It is easy to show that (\ref{Expand_NISTF2Gusfun_DiffScaPar_1}) is a
special case of the following expansion \cite[Eq.\
(6.12)]{Weniger/2007b}:
\begin{align}
  \label{ExpoPow2GLag}
  x^{\mu} \, \mathrm{e}^{u x} & \; = \; (1-u)^{-\alpha-\mu-1} \,
  \frac{\Gamma (\alpha+\mu+1)}{\Gamma (\alpha+1)}
  \notag \\
  & \qquad \times \, \sum_{n=0}^{\infty} \, {}_2 F_1 \left(-n,
    \alpha+\mu+1; \alpha+1; \frac{1}{1-u} \right) \, L_{n}^{(\alpha)} (x)
  \, ,
  \notag \\
  & \qquad \qquad \mu \in \mathbb{R} \setminus \mathbb{N}_{0} \, , \quad \Re
  (\mu+\alpha) > - 1 \, , \! \quad u \in (-\infty, 1/2) \, .
\end{align}
The condition $-\infty < u < 1/2$ is necessary to guarantee that this
expansion converges in the mean with respect to the norm of the weighted
Hilbert space $L^{2}_{\mathrm{e}^{-x} x^{\alpha}} (\mathbb{R}_{+})$. For
$u=0$, (\ref{ExpoPow2GLag}) simplifies to give (\ref{GenPow2GLag}).

If we insert the explicit expression (\ref{Def_GLagPol}) of the
generalized Laguerre polynomials into (\ref{ExpoPow2GLag}) and
interchange the order of summations, we also obtain a formal power series
in $x$. Unfortunately, an analysis of the resulting power series becomes
very difficult because of the terminating Gaussian hypergeometric series
${}_2 F_1$ in (\ref{ExpoPow2GLag}) (probably, an analysis of the behavior
of this ${}_2 F_1$ as $n \to \infty$ would be a nontrivial research
project in its own right). However, we can argue that the function
$z^{\mu} \exp (u z)$ with $\mu, u, z \in \mathbb{C}$ is only analytic at
$z=0$ in the sense of complex analysis if $\mu$ is a nonnegative integer,
$\mu=m$ with $m \in \mathbb{N}_0$, yielding $z^m \exp (u z) =
\sum_{n=0}^{\infty} u^n z^{m+n}/n!$. If $\mu$ is nonintegral, $\mu \in
\mathbb{C} \setminus \mathbb{N}_0$, a power series expansion of of
$z^{\mu} \exp (u z)$ around $z=0$ does not exist.

Thus, also for different scaling parameters $\beta \neq \gamma$, the
one-center limit $\bm{r}' = \bm{0}$ of the rearranged addition theorems
for $\chi_{N, L}^{M} (\beta, \bm{r} \pm \bm{r}')$ does not exist if the
principal quantum number $N$ is nonintegral, $N \in \mathbb{R} \setminus
\mathbb{N}$.

Apparently, Guseinov deliberately ignores even now the fact that the
Laguerre expansions (\ref{GenPow2GLag}) for $x^{\mu}$ and
(\ref{ExpoPow2GLag}) for $x^{\mu} \mathrm{e}^{u x}$ cannot be transformed
to power series expansions in $x$ if $\mu \in \mathbb{R} \setminus
\mathbb{N}_{0}$, although this had been emphasized in \cite[pp.\ 18 -
19]{Weniger/2007b}. In \cite[Eqs.\ (5) - (6)]{Guseinov/2007b}, Guseinov
expanded Slater-type functions $\chi_{N, L}^{M} (\beta, \mathbf{r})$ with
nonintegral principal quantum numbers as an infinite series of
Slater-type functions $\chi_{n, \ell}^{m} (\beta, \mathbf{r})$ with
integral principal quantum numbers, although these expansions do not
exist since their terms are for all but a finite number of indices
infinite. In \cite[Eqs.\ (7) - (10)]{Guseinov/2007b}, Guseinov tried to
resell essentially the same nonexisting expansion for what he calls
\emph{Coulomb-Yukawa like correlated interaction potentials}, which are
apart from a different normalization nothing but special Slater-type
functions.

From a mathematical point of view, a one-range addition theorem for a
function $f (\bm{r} \pm \bm{r}')$ is a mapping $\mathbb{R}^3 \times
\mathbb{R}^3 \to \mathbb{C}$. In my opinion, one-range addition theorems
have the highly advantageous feature that they provide a \emph{unique}
infinite series representation of $f (\bm{r} \pm \bm{r}')$ with
\emph{separated} variables $\bm{r}$ and $\bm{r}'$ that is valid for the
\emph{whole} argument set $\mathbb{R}^3 \times \mathbb{R}^3$. If we
accept this premise, then we have to conclude that Guseinov's
manipulations, which produced his rearranged addition theorems for
Slater-type functions $\chi_{N, L}^{M} (\beta, \bm{r} \pm \bm{r}')$, are
at least in the case of nonintegral principal quantum numbers $N \in
\mathbb{R} \setminus \mathbb{N}$ a complete failure.

The analysis of the one-center limits of rearranged one-range addition
theorems provides valuable insight in the case of Slater-type functions
with nonintegral principal quantum numbers, but it does not answer all
questions. In particular, my nonanalyticity argument allows no
conclusions about the validity of Guseinov's rearrangements in the case
of Slater-type functions with integral principal quantum numbers. Another
interesting but open question is whether Guseinov's rearrangements are in
the case nonintegral principal quantum numbers invalid for the whole
argument set $\mathbb{R}^{3} \times \mathbb{R}^{3}$ or whether they are
invalid only in the one-center limit. This is a practically very relevant
question. If only the one-center limit is invalid, then it would be
conceivable that Guseinov's rearranged one-range addition theorems might
be mathematically meaningful or possibly even numerically useful in a
restricted sense as approximations, although they do not exist for the
whole argument set $\mathbb{R}^3 \times \mathbb{R}^3$. This has to be
investigated.

These examples show that the situation is much more complicated than
originally anticipated by Guseinov. Obviously, a lot of work remains to
be done before we can claim with some confidence that we understand the
subtleties of Guseinov's rearrangements sufficiently well. It should also
be clear that the burden of proof lies in all cases with Guseinov.

Nevertheless, I do not think that it would be a good idea to invest too
much time and effort into an analysis of these most likely very difficult
open questions. In my opinion, it is simply a bad idea to construct
one-range addition theorems that use nonorthogonal functions as expansion
functions. It would be much better to focus on those one-range addition
theorems that are expansions in terms of complete and orthonormal
function sets.

The principal superiority of orthogonal expansion functions becomes
particularly evident in the case of one-range addition theorems of the
type of (\ref{CP_OnerangeAddThm_GusFun_k}) that do not converge in the
mean with respect to the norm of an appropriate Hilbert space, but only
weakly in the sense of generalized functions in suitably restricted
functionals.

For example, in Section \ref{Sec:DivOrthExp} I analyzed under which
conditions inner products $(f \vert g)$ with $f \notin \mathcal{H}$ and
$g \in \mathcal{H}$ are mathematically meaningful and whether the
divergent orthogonal expansion $f = \sum_{n=0}^{\infty} (\varphi_{n}
\vert f) \varphi_{n}$ nan be used in these inner products. I showed that
if $g$ belongs to the subset $\mathcal{F} \subset \mathcal{H}$ defined by
(\ref{Def_F}), then the expansion $(f \vert g) = \sum_{n=0}^{\infty} (f
\vert \varphi_{n}) (\varphi_{n} \vert g)$ converges if the expansion
coefficients $(f \vert \varphi_{n}) (\varphi_{n} \vert g)$ decay more
rapidly than $1/n$ as $n \to \infty$.

Ignoring all questions of convergence or existence, let us now assume
that both $f$ and $g$ can be expanded at least formally in terms of a
complete, but nonorthogonal function set $\{ \psi_n \}_{n=0}^{\infty}$:
\begin{align}
  f & \; = \; \sum_{m=0}^{\infty} \, F_{m}^{(\psi)} \, \psi_m \, ,
  \\
  g & \; = \; \sum_{n=0}^{\infty} \, G_{n}^{(\psi)} \, \psi_n \, .
\end{align}
If we insert these expansions into the inner product $(f \vert g)$, we
formally obtain the following double series:
\begin{equation}
   (f \vert g) \; = \; \sum_{m=0}^{\infty} \, \sum_{n=0}^{\infty} \,
   \bigl[ F_{m}^{(\psi)} \bigr]^{*} \, G_{n}^{(\psi)} \, 
   (\psi_m \vert \psi_n) \, . 
\end{equation}
There can be no doubt that it would be much harder to formulate
convergence criteria for this complicated double series than for the
comparatively simple series $(f \vert g) = \sum_{m=0}^{\infty} (f \vert
\varphi_{m}) (\varphi_{m} \vert g)$ which we obtain if we expand $f$ and
$g$ in terms of a complete and orthonormal function set $\{ \varphi_n
\}_{n=0}^{\infty}$.

This simple example shows that it is a highly dubious idea to expand
generalized functions in the sense of Schwartz \cite{Schwartz/1966a} in
terms of nonorthogonal function sets. At best, we would be confronted
with nontrivial technical problems.

\typeout{==> Section: Summary and Conclusions}
\section{Summary and Conclusions}
\label{Sec:SumConclu}

To some extent, divergent series are the dominant theme of this Reply. I
am fully aware that divergent series have been and to some extend still
are a highly controversial topic. There are still many who thoroughly
dislike divergent series and think that they should be banned from the
realm of rigorous mathematics. In their opinion, divergent series should
at best be considered to be some kind of mathematical pornography. In
addition, there are many others who -- either because of ignorance or
because of over-confidence -- wrongly believe that divergent series
cannot occur in their work and who thus tend to ignore questions of
convergence.

It is now widely accepted that divergent series play a very useful role.
They are indispensable tools in mathematics and in particular also in the
mathematical treatment of scientific problems.  Skeptical readers, who
still prefer to ignore divergent series, should search Google Scholar
(\texttt{http://scholar.google.com/}) for ``divergent series'' or for
related topics. They will be surprised by the large number of
applications of divergent series in different scientific disciplines.

There are principal differences between the divergent orthogonal
expansions considered in this Reply and the more familiar divergent power
series, which for instance occur abundantly in quantum mechanical
perturbation expansions or as asymptotic expansions for special
functions. Divergent power series can be used for the numerical
evaluation of the function they represent: With the help of suitable
summation techniques as for instance Borel summation, Pad\'{e}
approximants, or nonlinear sequence transformations it is frequently
possible to associate a finite value to a divergent power series.

In contrast, it is not intended to use the divergent orthogonal
expansions of this Reply for the direct numerical evaluation of the
function they represent. We only want to use these expansions in suitable
functionals -- typically multicenter integrals -- because we hope for
some formal simplifications. Actually, this applies to all one-range
addition theorems: They are only intermediate results which ultimately
produce series expansions for multicenter integrals.

The use of convergent expansions in integrals has the undeniable
advantage that normally only comparatively mild assumptions are needed to
guarantee that integration and summation can be interchanged and that the
resulting expansions converge. Nevertheless, the use of convergent
expansions in integrals is to some extent a luxury and not strictly
necessary. We are free to use a divergent expansion in an integral and
interchange integration and summation if we can guarantee that the
resulting expansion converges to the correct result.

Obviously, such an approach gives us additional possibilities, but it
would be naive to expect a free lunch: It is grossly negligent to use
divergent series in integrals without explicitly knowing criteria of
manageable complexity that guarantee the convergence of the resulting
expansions. This is probably the most serious flaw of Guseinov's work on
one-range addition theorems. Since he is apparently completely unaware of
the fact that divergent orthogonal expansions occur in his work, he has
no reason to think about additional criteria which could justify the use
of his divergent expansions in multicenter integrals.

In \cite[Abstract]{Guseinov/2007a}, Guseinov claims that all his formulas
were numerically tested, but this does not prove anything. A much more
profound understanding of these divergent expansions and their domains of
validity is needed, before they could be applied safely and in a
mathematically rigorous way. Otherwise, the use of Guseinov's divergent
one-range addition theorems in multicenter integrals would be purely
experimental.

Misconceptions about divergent series are also the core of Guseinov's
problems with his rearrangements of one-range addition theorems discussed
in Section \ref{Sec:RearrAddThm}. It seems that Guseinov wrongly believes
that it is always safe to rearrange the order of summations of double
series $\sum_{m=0}^{\infty} \sum_{n=0}^{\infty} a_{m, n}$.

In recent years, Guseinov and coworkers were able to publish a remarkably
large number of articles on one-range addition theorems
\cite{Guseinov/2001,Guseinov/2002b,Guseinov/2002c,Guseinov/2002d,%
  Guseinov/2003b,Guseinov/2003c,Guseinov/2003d,Guseinov/2003e,%
  Guseinov/2004a,Guseinov/2004b,Guseinov/2004c,Guseinov/2004d,%
  Guseinov/2004e,Guseinov/2004f,Guseinov/2004i,Guseinov/2004k,%
  Guseinov/2005a,Guseinov/2005b,Guseinov/2005c,Guseinov/2005d,%
  Guseinov/2005e,Guseinov/2005f,Guseinov/2005g,Guseinov/2006a,%
  Guseinov/2006b,Guseinov/Aydin/Mamedov/2003,Guseinov/Mamedov/2001c,%
  Guseinov/Mamedov/2002d,Guseinov/Mamedov/2003,Guseinov/Mamedov/2004b,%
  Guseinov/Mamedov/2004d,Guseinov/Mamedov/2004e,Guseinov/Mamedov/2004g,%
  Guseinov/Mamedov/2004h,Guseinov/Mamedov/2005c,Guseinov/Mamedov/2005d,%
  Guseinov/Mamedov/2005g,Guseinov/Mamedov/Rzaeva/2001,%
  Guseinov/Mamedov/Rzaeva/2002,Guseinov/Mamedov/Suenel/2002}. This fact
and the dubious quality of these articles raise obvious doubts about the
quality of our scientific publication system based on anonymous
peer-refereeing, and also on the competence of Guseinov's referees who
also failed to understand the mathematical subtleties of one-range
addition theorems. 

Of course, Guseinov disagreed with this conclusion which was first
expressed in \cite[p.\ 5]{Weniger/2007b}. In \cite[p.\
3]{Guseinov/2007a}, he stated:
\begin{quote}
  \sl The respectable referees very well understand and examined the
  published by Guseinov and his coworkers in the years from 1978 to 2006
  papers on one-range addition theorems.
\end{quote}
Unfortunately, I am not so optimistic. But it would be unfair to blame
exclusively Guseinov's referees. Refereeing Guseinov's manuscripts is
certainly not easy. It is Guseinov's trademark to produces a large number
of \emph{short} and largely \emph{overlapping} articles on essentially
the same topic. This makes it very hard even for a very competent referee
not to get lost in Guseinov's flood of publications and to keep track of
Guseinov's truly new results. Moreover, as I know from my own experience
as a referee, there is always the temptation to be less critical in the
case of a short manuscript than in the case of a (very) long manuscript.

In my opinion, part of the problem are \emph{short articles}. While there
can be no doubt that short articles are well suited to present new
experimental or computational results, they are basically unsuited for
predominantly theoretical or mathematical topics.

For example, in a theoretical article on multicenter integrals, it is
first necessary to provide a usually (very) long list of special
functions and other abbreviations and conventions. Then, it is necessary
to give a compact, but hopefully sufficiently comprehensive description
of the mathematical techniques, which are to be employed. To do these
things in a reasonable and for the reader beneficial way, we need at
least a few pages before we can start with the derivation and description
of new results.

If we nevertheless insist on writing (very) short articles, we can either
shrink the in my opinion very important introductory part to an absolute
minimum, or we can try to split the new results into numerous small
pieces. Either alternative is undesirable: If we choose the first
alternative, essential background information may be lacking and
readability will most likely suffer quite a bit, and if we choose the
latter alternative, we have to write numerous articles on essentially the
same topic that contain virtually nothing new. 

Of course, a compromise would also be possible: One could write a large
number of articles with highly condensed and thus more or less
incomprehensible introductory parts, that also present at best
infinitesimal increments of insight.

The problems with short articles can be demonstrated convincingly by
Guseinov's recent reprint \cite{Guseinov/2007c}. Guseinov's only new
result, which I could detect, are his Eqs.\ (6) - (7), which express a
Slater-type function $\chi_{N, L}^{M} (\beta, \mathbf{r})$ with
nonintegral principal quantum numbers as an infinite series of his
functions $\prescript{}{k}{\Psi}_{n, \ell}^{m} (\gamma, \bm{r})$ and
which corresponds to (\ref{Expand_NISTF2Gusfun_DiffScaPar_1}). In his
Eqs.\ (8) - (9), Guseinov tried to resell his Eqs.\ (6) - (7) as an
expansion for what he calls Coulomb-Yukawa like correlated interaction
potentials, which are apart from a different normalization nothing but
special Slater-type functions.

It is fairly easy to derive Guseinov's new expansion. We only have to
combine some well known properties of generalized Laguerre polynomials
with a formula from the book by Gradshteyn and Rhyzhik \cite[Eq.\
(7.414.7) on p.\ 850]{Gradshteyn/Rhyzhik/1994}. Of course, Guseinov's new
expansion can be published, but one may wonder whether Guseinov's new
series expansion alone justifies a new article. A change of the editorial
policy.of scientific journals with respect to short articles on
predominantly mathematical topics might be helpful.

\begin{appendix}
\typeout{==> Appendix: Terminology and definitions}
\section{Terminology and Definitions}
\label{App:Terminolgy}

For the set of \emph{positive} integers, I write $\mathbb{N} = \{ 1, 2,
3, \ldots \}$, and for the set of \emph{non-negative} integers, I write
$\mathbb{N}_0 = \{ 0, 1, 2, \ldots \}$. The real and complex numbers and
the set of three-dimensional vectors with real components are denoted by
$\mathbb{R}$, $\mathbb{C}$, and $\mathbb{R}^3$, respectively.
$\mathbb{R}_{+}$ is the set of real numbers $\ge 0$.

Fourier transformation is used in its symmetrical form, i.e., a function
$f \colon \mathbb{R}^3 \to \mathbb{C}$ and its Fourier transform
$\bar{f}$ are connected by the integrals
\begin{align}
  \label{Def_FT}
  \bar{f} (\bm{p}) & \; = \; (2\pi)^{-3/2} \int \,
  \mathrm{e}^{-\mathrm{i} \bm{p} \cdot \bm{r}} \, f (\bm{r})
  \, \mathrm{d}^3 \bm{r} \, ,
  \\
  \label{Def_InvFT}
  f (\bm{r}) & \; = \; (2\pi)^{-3/2} \int \,
  \mathrm{e}^{\mathrm{i} \bm{r} \cdot \bm{p}} \, \bar{f} (\bm{p}) 
  \, \mathrm{d}^3 \bm{p} \, ,
\end{align}

\typeout{==> Appendix: Hilbert Spaces}
\section{Hilbert Spaces}
\label{App:HilbertSpaces}

Let $\mathcal{V}$ be a vector space over the complex numbers $\mathbb{C}$
that possesses an inner product $( \cdot \vert \cdot ): \mathcal{V}
\times \mathcal{V} \to \mathbb{C}$, satisfying for all $u, v, w \in
\mathcal{V}$ and for all $\alpha \in \mathbb{C}$ \cite[p.\
36]{Reed/Simon/1980}
\begin{subequations}
  \label{Def_InnerProd}
\begin{align}
  ( u \vert u ) & \; \geq \; 0 \, , \\
  ( u \vert u ) & \; = \; 0 \; \Longleftrightarrow \; u \; = \; 0 \, , \\
  ( u \vert v + w ) & \; = \;
  ( u \vert v ) \, + \, ( u \vert w ) \, , \\
  ( u \vert \alpha v ) & \; = \; \alpha ( u \vert v ) \, , \\
  ( u \vert v ) & \; = \; ( v \vert u )^{*} \, .
\end{align}
\end{subequations}

Another essential concept is the norm $\Vert \cdot \Vert: \mathcal{V} \to
\mathbb{R}$ of the elements of a vector space $\mathcal{V}$, satisfying
for all $u, v \in \mathcal{V}$ and for all $\alpha \in \mathbb{C}$
\cite[p.\ 8]{Reed/Simon/1980}
\begin{subequations}
  \label{Def_Norm}
\begin{align}
  \Vert u \Vert & \; \geq \; 0 \, , \\
  \Vert u \Vert & \; = \; 0 \; \Longleftrightarrow \; u \; = \; 0 \, , \\
  \Vert \alpha u \Vert & \; = \;
  \vert \alpha \vert \, \Vert u \Vert \, , \\
  \Vert u+v \Vert & \; \le \; \Vert u \Vert \, + \, \Vert v \Vert \, .
\end{align}
\end{subequations}
Obviously, $\Vert u \Vert = \sqrt{(u \vert u)}$ with $u \in \mathcal{V}$
is a norm satisfying these conditions.

A vector space $\mathcal{V}$ over the complex numbers $\mathbb{C}$ is
called a Hilbert space, if it possesses an inner product $( \cdot \vert
\cdot )$ satisfying (\ref{Def_InnerProd}), and if $\mathcal{V}$ is
complete with respect to the norm defined by $\Vert u \Vert = \sqrt{(u
  \vert u)}$. Completeness implies that every Cauchy sequence in
$\mathcal{V}$ converges with respect to this norm to an element of
$\mathcal{V}$.

In bound-state electronic structure calculations, we have to take into
account Born's statistical interpretation of the wave function. Thus, an
obvious \emph{inner product} for functions $f, g \colon \mathbb{R}^3 \to
\mathbb{C}$, that can be used as basis functions in atomic and molecular
electronic structure calculations, can be defined according to
\begin{equation}
  \label{InnerProd}
( f \vert g )_2 \; = \;
\int \bigl[ f (\bm{r}) \bigr]^{*} \, g (\bm{r}) \,
\mathrm{d}^3 \bm{r} \, .
\end{equation}
As usual, the integration extends over the whole $\mathbb{R}^3$.

On the basis of the inner product (\ref{InnerProd}), the \emph{norm} of
a function $f \colon \mathbb{R}^3 \to \mathbb{C}$ is defined according to
\begin{equation}
  \label{Norm_2}
\Vert f \Vert_2 \; = \; \sqrt{( f \vert f )_2} \, .
\end{equation}

The \emph{Hilbert} space $L^{2} (\mathbb{R}^3)$ of \emph{square
  integrable} functions is defined via the norm (\ref{Norm_2}) according
to
\begin{align}
  \label{HilbertL^2}
  L^{2} (\mathbb{R}^3) & \; = \; \Bigl\{ f \colon \mathbb{R}^3 \to
  \mathbb{C} \Bigm\vert \, \int \, \vert f (\bm{r}) \vert^2 \,
  \mathrm{d}^3 \bm{r} < \infty \Bigr\} 
  \notag \\[1\jot]
  & \; = \; \bigl\{ f \colon \mathbb{R}^3 \to \mathbb{C} \big\vert \,
  \Vert f \Vert_2 < \infty \bigr\} \, .
\end{align}

The formalism of Hilbert spaces can be generalized to include weight
functions.  If $w \colon \mathbb{R}^3 \to \mathbb{R}_{+}$ is a suitable
\emph{positive} weight function, we define the \emph{inner product} with
respect to the weight function $w$ for functions $f, g \colon
\mathbb{R}^3 \to \mathbb{C}$ according to
\begin{equation}
  \label{InnerProd_w}
( f \vert g )_{w, 2} \; = \;
\int \bigl[ f (\bm{r}) \bigr]^{*} \, w (\bm{r}) \,
g (\bm{r}) \, \mathrm{d}^3 \bm{r} \, .
\end{equation}
As in (\ref{InnerProd}), the integration extends over the whole
$\mathbb{R}^3$. It is easy to show that $( f \vert g )_{w, 2}$ with $w
(\bm{r}) \ge 0$ is indeed an inner product satisfyng
(\ref{Def_InnerProd}).

On the basis of the inner product (\ref{InnerProd_w}), the \emph{norm}
of a function $f \colon \mathbb{R}^3 \to \mathbb{C}$ with respect to the
weight function $w$ is defined according to
\begin{equation}
  \label{Norm_w_2}
\Vert f \Vert_{w, 2} \; = \; \sqrt{( f \vert f )_{w, 2}} \, .
\end{equation}

The \emph{Hilbert} space $L_{w}^{2} (\mathbb{R}^3)$ of \emph{square
  integrable} functions with respect to the weight function $w$ is
defined via the norm (\ref{Norm_w_2}) according to
\begin{align}
  \label{HilbertL_w^2}
  L_{w}^{2} (\mathbb{R}^3) & \; = \; \Bigl\{ f \colon \mathbb{R}^3 \to
  \mathbb{C} \Bigm\vert \, \int \, w (\bm{r}) \, \vert f (\bm{r}) \vert^2
  \, \mathrm{d}^3 \bm{r} < \infty \Bigr\} 
  \notag \\[1\jot]
  & \; = \; \bigl\{ f \colon \mathbb{R}^3 \to \mathbb{C} \big\vert \,
  \Vert f \Vert_{w, 2} < \infty \bigr\} \, .
\end{align}

It is not necessary to use the coordinate representation for the
definition of the Hilbert spaces $L^{2} (\mathbb{R}^3)$. Instead, the
momentum representation can also be used. This is a consequence of the
well-known fact that Fourier transformation defined via (\ref{Def_FT})
and (\ref{Def_InvFT}) maps $L^{2} (\mathbb{R}^3)$ onto $L^{2}
(\mathbb{R}^3)$ in a one-to-one manner such that inner products are
conserved \cite[Theorem IX.6 on p.\ 10]{Reed/Simon/1975}. Thus, $f, g \in
L^{2} (\mathbb{R}^3)$ implies that the Fourier transforms $\bar{f}
(\bm{p})$ and $\bar{g} (\bm{p})$ are also elements of $L^{2}
(\mathbb{R}^3)$. In addition, the inner product (\ref{InnerProd})
satisfies
\begin{equation}
  \label{InnerProd_FT}
( f \vert g )_2 \; = \; \int \bigl[ \bar{f} (\bm{p}) \bigr]^{*} \,
\bar{g} (\bm{p}) \, \mathrm{d}^3 \bm{p} \, .
\end{equation}

\typeout{==> Appendix: Laguerre polynomials}
\section{Laguerre polynomials}
\label{App:LagPol}

The generalized Laguerre polynomials $L_{n}^{(\alpha)} (x)$ possess the
following explicit expressions \cite[p.\
240]{Magnus/Oberhettinger/Soni/1966}:
\begin{subequations}
  \label{Def_GLagPol}
  \begin{align}
    \label{GLag_FinSum}
    L_{n}^{(\alpha)} (x) & \; = \; \sum_{\nu=0}^{n} \, (-1)^{\nu} \,
    \binom{n+\alpha}{n-\nu} \, \frac{x^{\nu}}{\nu!}
    \\
    \label{GLag_1F1}
    & \; = \; \frac{(\alpha+1)_n}{n!} \, {}_1 F_1 (-n; \alpha+1; x) \, .
  \end{align}
\end{subequations}

The generalized Laguerre polynomials satisfy for $\Re (\alpha) > - 1$
and $m, n \in \mathbb{N}_0$ the following orthogonality relationship
\cite[p.\ 241]{Magnus/Oberhettinger/Soni/1966}:
\begin{equation}
  \label{GLag_Orthogonality}
\int_{0}^{\infty} \, x^{\alpha} \, \mathrm{e}^{-x} \,
L_{m}^{(\alpha)} (x) \, L_{n}^{(\alpha)} (x) \, \mathrm{d} x \; = \;
\frac{\Gamma (\alpha+n+1)}{n!} \, \delta_{m n} \, .
\end{equation}

The completeness of the generalized Laguerre polynomials in the weighted
Hilbert space
\begin{align}
  \label{HilbertL^2_Lag}
  & L^{2}_{\mathrm{e}^{-x} x^{\alpha}} (\mathbb{R}_{+}) \; = \;
  \notag \\
  & \; = \; \Bigl\{ f \colon \mathbb{R}_{+} \to \mathbb{C} \Bigm\vert \,
  \int_{0}^{\infty} \,\mathrm{e}^{-x} \, x^{\alpha} \, \vert f (x)
  \vert^2 \, \mathrm{d} x < \infty \Bigr\}
\end{align}
is a classic result of mathematical analysis (see for example \cite[p.\
33]{Higgins/1977}, \cite[pp.\ 349 - 351]{Sansone/1977}, or \cite[pp.\ 235
- 238]{Tricomi/1970}).

\typeout{==> Appendix: Guseinov's Function Set}
\section{Guseinov's Function Set}
\label{App:GusFun}

In \cite[Eq.\ (1)]{Guseinov/2002b}, Guseinov introduced a fairly large
class of complete and orthonormal functions which can be expressed
as follows:
\begin{equation}
  \label{Def_Psi_Guseinov}
  \prescript{}{k}{\Psi}_{n, \ell}^{m} (\beta, \bm{r}) \; = \; \left[
  \frac{(2\beta)^{k+3} (n-\ell-1)!}{(n+\ell+k+1)!} \right]^{1/2} \, 
  \mathrm{e}^{-\beta r} \, L_{n-\ell-1}^{(2\ell+k+2)} (2 \beta r) \,
  \mathcal{Y}_{\ell}^{m} (2 \beta \bm{r}) \, .
\end{equation}
The indices satisfy $n \in \mathbb{N}$, $k = -1, 0, 1, 2, \dots$, $\ell
\in \mathbb{N}_0 \le n - 1$, $-\ell \le m \le \ell$, and the scaling
parameter $\beta$ is positive.

Guseinov's functions satisfy the orthonormality relationship (compare
also \cite[Eq.\ (4)]{Guseinov/2002c})
\begin{equation}
  \label{Psi_Guseinov_OrthoNor}
  \int \, \bigl[ \prescript{}{k}{\Psi}_{n, \ell}^{m} (\beta, \bm{r})
  \bigr]^{*} \, r^k \, \prescript{}{k}{\Psi}_{n', \ell'}^{m'} (\beta,
  \bm{r}) \, \mathrm{d}^3 \bm{r} \; = \; 
  \delta_{n n'} \, \delta_{\ell \ell'} \, \delta_{m m'} \, .
\end{equation}
Accordingly, Guseinov's functions are a complete and orthonormal set in
the weighted Hilbert space $L_{r^k}^{2} (\mathbb{R}^3)$ with $k = -1, 0,
1, 2, \dots$, which is defined via the inner product
\begin{equation}
  \label{InnerProd_r^k}
  ( f \vert g )_{r^k, 2} \; = \;
  \int \bigl[ f (\bm{r}) \bigr]^{*} \, r^k \,
  g (\bm{r}) \, \mathrm{d}^3 \bm{r}
\end{equation}
and the norm
\begin{equation}
  \label{Norm_r^k_2}
\Vert f \Vert_{r^{k}, 2} \; = \; \sqrt{( f \vert g )_{r^k, 2}}
\end{equation}
according to
\begin{align}
  \label{HilbertL_r^k^2}
  L_{r^k}^{2} (\mathbb{R}^3) & \; = \; \Bigl\{ f \colon \mathbb{R}^3 \to
  \mathbb{C} \Bigm\vert \, \int \, r^k \, \vert f (\bm{r}) \vert^2 \,
  \mathrm{d}^3 \bm{r} < \infty \Bigr\} \notag \\[1\jot]
  & \; = \; \bigl\{ f \colon \mathbb{R}^3 \to \mathbb{C} \big\vert \,
  \Vert f \Vert_{r^k, 2} < \infty \bigr\} \, .
\end{align}

\typeout{==> Appendix: Expansion of the Yukawa Potential}
\section{Expansion of the Yukawa Potential}
\label{App:ExpYukawa2GusFun}

It is relatively easy to construct an expansion of the Yukawa potential
in terms of Guseinov's function $\prescript{}{k}{\Psi}_{n, \ell}^{m}
(\beta, \bm{r})$. The Yukawa potential is a special Slater-type function
satisfying
\begin{equation}
\frac{\mathrm{e}^{-\beta r}}{r} \; = \; (4\pi)^{1/2} \, \beta \,
\chi_{0, 0}^{0} (\beta, \mathbf{r}) \, .
\end{equation}
Setting $N=L=M=0$ in (\ref{NISTF2Gusfun_EqScaPar}) yields:
\begin{equation}
  \label{Yukawa2Gusfun_EqScaPar}
  \frac{\mathrm{e}^{-\beta r}}{r} \; = \; 
  \bigl[ 2\pi/(2\beta)^{k+1} \bigr]^{1/2} \, 
  \Gamma (k+2) \, \sum_{\nu=0}^{\infty} \,
  \frac{{\nu}!}{\bigl[ (\nu+k+2)! \, \nu! \bigr]^{1/2}}
  \, \prescript{}{k}{\Psi}_{\nu+1, 0}^{0} (\beta, \bm{r}) \, .  
\end{equation}
This expansion converges in the mean with respect to the norm
(\ref{Norm_r^k_2}) of the weighted Hilbert space $L_{r^k}^{2}
(\mathbb{R}^3)$ with $k = -1, 0, 1, 2, \dots$ if the squares of the
coefficients on the right-hand side decay more rapidly than $1/\nu$ as
$\nu \to \infty$.

The behavior of the coefficients in (\ref{Yukawa2Gusfun_EqScaPar}) can be
analyzed with the help of the following asymptotic expression for the
ratio of two gamma functions \cite[Eq.\ (6.1.47) on p.\
257]{Abramowitz/Stegun/1972}:
\begin{equation}
  \label{AsyGammaRatio}
  \frac{\Gamma(z+a)}{\Gamma(z+b)} \; = \; 
  z^{a-b} \, + \, \mathrm{O} \bigl(z^{a-b-1}\bigr) \, , 
  \qquad z \to \infty \, .
\end{equation}
We then obtain the following asymptotic estimates for the $\nu$-dependent
part of the square of the coefficients on the right-hand side of
(\ref{Yukawa2Gusfun_EqScaPar}):
\begin{equation}
\left[ \frac{{\nu}!}{\bigl[ (\nu+k+2)! \, \nu! \bigr]^{1/2}} \right]^{2}
\; = \; \frac{{\nu}!}{(\nu+k+2)!} \; = \; 
\nu^{-k-2} + \mathrm{O} \bigl(\nu^{-k-3}\bigr) \, ,
\qquad \nu \to \infty \, .
\end{equation}
Thus, the expansion (\ref{Yukawa2Gusfun_EqScaPar}) converges in the mean
with respect to the norm (\ref{Norm_r^k_2}) of the weighted Hilbert space
$L_{r^k}^{2} (\mathbb{R}^3)$ for $k = 0, 1, 2 \dots$ and diverges for
$k=-1$. This is in agreement with the fact that the Yukawa potential
belongs to the weighted Hilbert space $L_{r^k}^{2} (\mathbb{R}^3)$ for $k
= 0, 1, 2 \dots$, but not for $k=-1$.
\end{appendix}

%
%
%
{\small

\providecommand{\SortNoop}[1]{} \providecommand{\OneLetter}[1]{#1}
  \providecommand{\SwapArgs}[2]{#2#1}

}
%


\end{document}